\documentclass[letterpaper]{article}

\usepackage{parskip}
\usepackage{graphicx}
\usepackage{amsmath}
\usepackage{caption}
\usepackage{chngpage}
\usepackage{amssymb}
\usepackage{textgreek}
\usepackage [english]{babel}
\usepackage [autostyle, english = american]{csquotes}
\MakeOuterQuote{"}

\begin{document}
\title{Are Rents Excessive in the Central City?:\\ A Geospatial Analysis}
\author{Scott W. Hegerty\\
Department of Economics\\
Northeastern Illinois University\\
Chicago, IL 60625\\S-Hegerty@neiu.edu
}
\date{\today}
\maketitle

\begin{abstract}
In many U.S. central cities, property values are relatively low, while rents are closer to those in better-off neighborhoods. This gap can lead to relatively large profits for landlords, and has been referred to as "exploitaton" for renters. While much of this gap might be explained by risk, factors such as income and race might play important roles as well. This study calculates Census tract-level measures of the rent-to-property-value (\textit{RPV}) ratio for 30 large cities and their surrounding metropolitan areas. After examining the spatial distribution of this ratio and relationships with other socioeconomic variables for Milwaukee and three other cities, Z-scores and quantiles are used to identify "extreme" \textit{RPV} values nationwide. "Rust Belt" cities such as Detroit, Cleveland, and Milwaukee are shown to have higher median and 95\% values than do West Coast cities such as Seattle and San Francisco. A spatial lag regression estimation shows that, controlling for income, property values, and vacancy rates, racial characteristics often have the "opposite" signs from what might be expected and that there is little evidence of purely race-based "exploitation" of renters. A significantly negative coefficient for the percentage of Black residents, for example, might suggest that the \textit{RPV} ratio is lower in a given tract, all else equal. While this study shows where \textit{RPV} values are hghest within as well as between cities, further investigation might uncover the drivers of these spatial differences more fully.    
\end{abstract}

\section{Introduction}
Matthew Desmond’s 2016 book \textit{Evicted: Poverty and Profit in the American City} helped bring attention to the experiences of renters in central-city Milwaukee, who often pay large shares of their incomes for substandard housing units that are often poorly maintained. The rents paid, while lower than in more affluent areas, are not proportional to the low property values typical of these neighborhoods. Anecdotally and through news reports and other non-academic channels\footnote{One intereresting article, by Cary Spivak, ran in the Milwaukee Journal Sentinel (April 15, 2021) and is titled "Out-of-state corporate landlords are gobbling up Milwaukee homes to rent out, and it's changing the fabric of some neighborhoods."}, there have been reports of landlords coming from outside the area to purchase rental properties cheaply, and then profit from this difference between rents and property values (particularly if they minimize maintenance). This contributes to a situation where a lack of community ties and the quest for profit collide with the economic needs of residents with nowhere else to turn. 

This phenomenon is not confined to any single city; Chicago has its own areas where renters pay large shares of their incomes simply for the right not to be thrown out. While a handful of studies have  examined the rent burdens faced by urban residents, no study to date has focused specifically on geographic variation within U.S. central cities and the unique conditions experienced in these neighborhoods.  This study does so, examining the rent-to-property-value (\textit{RPV}) ratio at the Census tract level for 30 large U.S. cities. 

The study of differences in rents, particularly among Whites and Blacks, received particular attention duing the 1960s, when trends in racial segregation were increasing. Haugen and Heins (1969) examine the ratio of median rents between the two racial groups in the context of "white flight." In highly segregated urban neighborhoods, Blacks paid higher rent than Whites, particularly where there was a large increase in the non-White proportion of the population. 

Since then, a number of studies have analyzed property values and rents separately. In the first group, Grether and Mieszkowski (1974) apply a heedonic pricing model, which includes a number of idiosyncratic, property-specific variables, to New Haven, Connecticut. Structural and neighborhood characteristics explain a large share of these differences. More recently, Ding and Knapp (2003) examine Cleveland, finding that outmigration has had a negative eeffect on property values. Examining sales prices in Florida, Ihlanfeldt and Mayock (2009) find evidence of discrimination against Black and Asian residents. Using data from more than two million home sales, Bayer et al. (2017) conclude that Black and Hispanic residents paid more for housing, all other factors equal.

The second group of studies focuses on the determinants of rent in urban neighborhoods. Gilderbloom and Applebaum (1987) treat rental markets as noncompetitive, focusing on landlord behavior. Analyzing 140 urban housing markets, the authors show that income and median  house prices affected rents, but that the vacancy and rental rates are not significant. In contrast to the study cited above, the percentage of non-White residents has a significantly negative coefficient in only one model specification. Zeitz and Sirmans (2004) analyze the "rent gap" as part of a literature review on a number of topics, and Early et al. (2018) estimate a hedonic pricing model and find that Black residents tend to pay a premium only in mainly White areas. Colburn and Allen (2018) focus on "rent burden" proportion to income and find that financial stress remained high after the Great Recession.

Tying these two concepts together, Desmond and Wilmers (2019) explicitly calculate the ratio of rents to property values. They cite maintenance costs (particularly for older homes) and risk (both non-payment of rent and turnover among renters) as possible reasons why rents do not fall proportionally with property values. Their ratio is called "exploitation," but no mention is made of any "baseline" value that might be exceeded in certain neighborhoods. The study includes detailed expense information, to study the costs of property ownership that can reduce profits, and the Milwaukee Area Renters Study is also used as a supplement to the main study. Median regression models applied to Census block groups in Milwaukee and nationally include poverty rates and the percentages of Black residents, as well as controls such as the number of units per building (to capture returns to scale). The authors find that all three of these variables have positive effects on the rent/property value ratio at both scales of analysis. The study also examines profit, finding that says maintenance costs are low. A quantile regression estimation for Milwaukee shows that in the poorest neighborhoods, losses to landlords are high. This finding highlights the role of risk in explaining high rent/property value ratios.

The current study extends this analysis of this ratio to incorporate spatial characteristics, both in terms of mapping census tracts for major U.S. cities and by conducting spatial regression analyses for each city and MSA. Overall, while poor neighborhoods and those with large Black populations tend to have high \textit{RPV} ratios, controlling for income, property values, and vacancy rates does not show that White neighborhoods suffer less exploitation. While these findings warrant further study, there is evidence that "Rust Belt" cities tend to have the highest median and "extreme" values than other cities in this analysis.

The paper proceeds as follows. Section 2 explains the underlying theory and the methodology, and Section 3 provides the results. Section 4 concludes.

\section{Theory and Methodology}
According to financial theory, the price of any asset can be calculated as the sum of discounted expected future dividend payments. A share of stock, for example, will be worth more today if these payments are higher. But since the future is less tangible, future payments are worth less in today's terms. The sum of future payments--even if on never sells the stock--are therefore not infinite. 

Any change in "patience," which is a shift in relative importance between the present and future, will change the present value of later payments. Risk also has a major impact on expected payments; if an investor suspects that they might not get paid back, this will push the stock price down immediately. While these principles are often applied to finance, they also hold in property markets. 

Buyers calculate the future stream of rent payments, but also keep in mind the odds of a unit sitting vacant between renters, or the loss of income if a renter is unable or unwilling to pay regularly. Oftentimes, occasional missed rent payments are preferred to having a vacant unit, but either might lead to higher rents overall. Maintenance costs also reduce landlords' profits, and might be higher in older, central-city neighborhoods. These costs, like risk, might be passed on to the renter. Understanding the relationship of payments and property value, as well as neighborhood characteristics and risk, is key to the current study.  
 
Thus study uses tract-level data from the 2019 American Community Survey (ACS 5-year estimates) for 30 large U.S. cities and their metropolitan areas. The main variable of interest is the \textit{RPV} ratio. Based on the availability of data on rent, \textit{RPV} is calculated as Median Gross Rent - 2 Bedrooms divided by Median Value (dollars). This is annualized by multiplying by 12:

\begin{equation}
RPV = 12\times MedRent/MedVal
\end{equation}

As noted above, this variable is similar to the "exploitation" measure of Desmond and Wilmers (2019). Their simple measure does not estimate any type of "excess" rent; here, extreme values are captured using z-scores for \textit{RPV}. A further analysis could estimate a predictive model and use its residuals to capture "excess" \textit{RPV}, for example. The stastistical properties of the unadjusted \textit{RPV} measure are examined, while the adjusted \textit{ZRPV} is the main dependent variable in the econometric analysis. 

Other variables used in this study are also calculated using 2019 ACS 5-year estimates. These include tract-level median income, the vacancy rate (to proxy risk), the rental rate, and the rental vacancy rate. Racial characteristics (the percentages White and Black) are also entered in the model, with an additional focus placed on tracts with a population that is greater than 50\% Black. All measures are calculated for City tracts and for MSA tracts, using Census-defined counties to determine the metropolitan areas. 

The four "test case" cities of Milwaukee, Chicago, Atlanta, and San Francisco are mapped, and bivariate relationships between tract-level \textit{RPV} and a set of socioeconomic variables are plotted. Majority-Black tracts are highlighted in this analysis as well. Citywide correlations are then calculated between \textit{RPV} and a number of socioeconomic variables for all 30 cities, and a broader set of correlations are presented for the city of Milwaukee. Citywide median \textit{RPV} values are plotted against a number of variables  for 30 cities. This process is repeated for the MSA level as well, to capture any city-suburban differences. 

Next, the quantiles for \textit{RPV} values are calculated for all cities (and their MSAs). In addition to presenting overall statistics on these measures, the cities are ranked by median \textit{RPV} value, as well as the corresponding 50\%, 75\%, 90\%, and 95\% quantiles are plotted. A clear pattern emerges, with cities such as Detroit, Cleveland, and Milwaukee on one end, and Seattle and San Francisco on the other. The same procedure is applied to MSA-level data; differences in order show how core cities differ from their surrounding metropolitan areas in many, but not all, cases.

Finally, the determinants of variation in each city's or MSA's \textit{ZRPV} are analyzed using spatial lag regression methods. The underlying model places \textit{ZRPV} as a function of (log) median income (\textit{lnMedY}), the areawide (city or MSA) median property value (\textit{lnMedVal}), the vacancy rate (\textit{PERCVAC}), the rental rate (\textit{PERCRENT}), and a measure of racial characteristics. Because segregation often leads to a large negative correlation between the two, \textit{PERCWHT} and \textit{PERCBLK} are entered into separate specifications.
The regression equation can be written as: 

\begin{equation}
y = (I-\rho )^{-1} (X\beta +\epsilon )
\end{equation}

where \textit{y} = \textit{ZRPV}, \textit{X} is a matrix of explanatory variables described above, $\rho$ measures spatial autocorrelation, \textit{I} is an identity matrix, and \textit{W} is a spatial weights matrix. Following the advice of LeSage (2014), this is kept as simple as possible; here, it is a Queen contiguity matrix of order one. The results of this model will help show the roles of race and income in determining "excess" rent-to-property-value ratios, even when controlling for overall property values, vacancies and risk, and rental characteristics. The results for this model and other modes of analysis are presented below.

\section{Results}
Figure 1 shows the tract-level \textit{RPV} ratio for Milwaukee. This measure is highest in tracts that are lower-income and less White or more Black, primarily on the North and near South Sides. There are strong negative relationships between \textit{RPV} and income and property values, with a clear concentration of majority-Black tracts among those with the lowest mean property values. While low-income majority Black tracts can be found along the entire range of \textit{RPV} values, they tend to be clustered above the median value. This suggests that there are disproportionately high rent burdens in these areas. The tracts with the highest \textit{RPV} values--exceeding the 95\% quantile value--are primarily majority Black. There is a weak but positive relationship between property values and rents, supporting the idea that rents do not tend to fall proportionally in central-city neighborhoods.

Figure 2 shows similar patterns for Chicago, where the \textit{RPV} ratio is highest in a number of majority-Black Census tracts, which are often located on the South and West sides. The main difference with Milwaukee is the stronger relationship between rents and property values, perhaps because of Chicago's larger number of wealthy neighborhoods.

Atlanta, depicted in Figure 3, has a clear deliniation between Black neighborhoods with higher \textit{RPV}s, and White neighborhoods with low ratios. Incomes, rents, and property values register the same split as well. For this city, the strong negative relationship between median rents and \textit{RPV} suggests that high rents, rather than low property values, drive the gap more in Atlanta than in Milwaukee or Chicago.  \textit{RPV}'s (negative) relationship with median property values is stronger than is the (relatively flat) connection with income or its negative relationship with rent. There is also a somewhat weak, but still positive, relationship between median rents and property values. San Francisco, presented in Figure 4, represents an entirely different type of city. The only two majority-Black tracts have relatively low \textit{RPV} ratios.

Some general summary statistics for all 30 core cities are presented in Table 1. As might be expected, poverty rates and median incomes are negatively related with one another throughout the sample. Rental vacancy rates, while overall relatively low, are highest in Southern cities such as Houston and Jacksonville; in general, they are not closely related to cities' \textit{RPV} values. Median rents are not drastically different at the MSA level; summary statistics for these are given in Table 2. Property values increase in some cases (such as Raleigh, NC), but decrease in many others--likely because more semi-rural areas are included in the wider areas. In some cases, such as Atlanta, median rents are almost identical between the two units of analysis.

The specific case of Milwaukee is singled out for detailed correlation analysis in Table 3. Low incomes and high poverty rates are highly correlated with \textit{RPV} ratios in the city; high ratios are more closely connected to low median property values than to high rents in this case. Housing tenure (the percentage of renters) and the rental vacancy rate exhibit weaker connections to \textit{RPV}. Among the other variables, the rental rate and the rental vacancy rate are positively correlated, and as may be expected, the percentages of Black and White residents across tracts have a strong negative relationship. These relationships do not necessarily hold for other cities, however, particularly for those in the Southern and Western United States.

Correlations between \textit{RPV} and the explanatory variables are presented in Table 4. Indeed, the coefficients are quite diverse. While the correlations between \textit{RPV} and median income and the poverty rate are opposite in sign in nearly all cases, they differ by city and by region.  Poverty rates and the \textit{RPV} ratio have strong positive correlations in cities such as Atlanta, Buffalo, Jacksonville, and Philadelphia; the percentage of tracts' Black populations are also closely connected to \textit{RPV} in these cities. There is almost no relationship between \textit{RPV} and poverty or racial characteristics in Denver, on the other hand, and there is a negative correlation between \textit{RPV} and poverty for Boston and San Francisco. For these latter two cities, \textit{RPV} is positively related to median rents.  
Vacancy rates have both positive and negative correlations with \textit{RPV}, with no clear pattern, while overall relationships between \textit{RPV} and the rental vacancy rate, as well as median property age (captured here as the log of the median year built), are rather weak.

Table 5 provides some "baseline" numbers for the \textit{RPV} ratio. While Detroit's median value of 23.6 is a true outlier, cities such as Baltimore and Milwaukee have values above 10. Many cities (such as Chicago and Houston) have median \textit{RPV}s between 5 and 10, with Seattle's and San Francisco's below 5. MSA \textit{RPV} values are often lower than core-city ratios (which is the case for Austin, Detroit, and Milwaukee), but they can also be roughly the same (Atlanta and San Diego), or even lower (Boston, Miami, Washington). A visual depiction might be more informative.
 
Figure 5 shows scatterplots of each citywide median \textit{RPV} value versus the median values of other key socioeconomic variables. From these, one can infer both the general relationships and the presence of any outliers. There are clear positive relationships between \textit{RPV} and the poverty rate, the vacancy rate, the rental vacancy rate, and the percentage of Black residents, with negative relationships between \textit{RPV} and median income, the percentage White, median rents, and median property values. Detroit, and to a lesser extent Cleveland, stand out in nearly all the graphs, but there are relatively few true "outliers." One exception is with regard to the rental vacancy rate, where "Rust Belt" cities such as Milwaukee and Detroit have higher ratios than this variable would suggest. Orlando's large upper quantiles might warrant further investigation.

There are important differences at the MSA level, which is depicted in  Figure 6.  Cities with higher median incomes tend to have lower \textit{RPV} ratios, for example, but Washington and Boston have higher than expected values. The Atlanta metro has a lower ratio than would be predicted solely by its percentages of Black residents; the Thiel-Sen (1968) regression lines are "flatter" at the MSA level in general. "Rust Belt" cities such as Buffalo still have higher \textit{RPV} ratios than their vacancy rates would predict, but overall, more vacancies is associated with higher ratios. Lower property values are associated with higher \textit{RPVs} as well, even outside the core cities.

The quantile values of each city's \textit{RPV} ratio are shown in Figure 7. There is a clear pattern in "Rust Belt" cities such as Detroit, Cleveland, Milwaukee, Pittsburgh, and St. Louis, with the highest median values as well as high extreme (90\% and 95\%) values. At the other extreme lie wealthy West Coast cities such as Seattle and San Francisco, as well as the two largest U.S. cities. Atlanta and Chicago lie near the middle of the distribution. In the U.S. South, Houston and Jacksonville have relatively high medians; the only "Northern" city with a low ranking is wealthy and well-educated Boston. 

Finally, the results of the spatial regressions are provided for core cities and MSAs in Tables 6 and 7, respectively. For the former, median property values have a significantly negative relationship with \textit{ZRPV} for all 30 cities, suggesting that high ratios are often driven by the denominator while rents in the numerator are more stable. Median income has a significantly positive association with extreme values of this ratio--with the exception of the low t-values in the cases of Buffalo, Columbus, Pittsburgh, and St. Louis. Perhaps this is indicative of unique effects in the "Rust Belt" that were shown above. The percentage of renters has a significantly positive coefficient in the majority of cases, although cities such as Buffalo require further attention due to their low t-values. The vacancy rate, which proxies risk as well as neighborhood conditions, has a significantly positive effect in only 18 of the 30 cities. The coefficients are significant for Detroit, Milwaukee, and Pittsburgh--but also Phoenix, and not Atlanta or Cleveland--making any easy conclusions regarding city type difficult. 

Likewise, racial characteristics are only significant in half of the cities. The percentage White has a positive association with \textit{ZRPV} for 16 cities, including Milwaukee, Atlanta, Washington, Seattle, and San Francisco. The percentage Black has a negative correlation in 14 cities. The specifications are "matched" (positive-negative or insignificant-insignificant) in 26 cases; while for Chicago, only the percentage Black has a significant coefficient, the cities of Charlotte, Milwaukee, and Los Angeles have significant coefficients only in the specification that includes the percentage White. While these results are not definitively conclusive, they do suggest that there is little evidence that majority-Black tracts have \textit{higher} rent-to-property-value ratios than do majority-White tracts, when controlling for other factors. But low-income areas with low property values still need to be addressed.

Turning to the MSA-level estimation, the control variables have an even clearer effect. Median income has a significantly positive coefficient, and median property values have significantly neative effects, in all cases. The percentage of renters has a significantly positive impact on \textit{ZRPV} in all cities except Buffalo. The effect of race is now more mixed in sign, however: seven cities have significantly positive coefficients for the percentage White, but four (Buffalo, Nashville, Pittsburgh, and St. Louis) have negative coefficients. The suburbs of these MSAs might have relatively high degrees of segregation. Nine cities have significantly negative cofficients on the percentage Black, but nine (including Chicago, Milwaukee, and St. Louis, have significantly positive effects. This suggests that suburban segregation might be of particular importance in these areas. Further research would be necessary to uncover the underlying drivers of this phenomenon, however.

Overall, the spatial distributions of the \textit{RPV} ratios within (and between) cities, as well as the spatial regression analysis conducted here, point to three key findings. First, there is a class of cities--often located in the "Rust Belt"--where the median and "extreme" \textit{RPV} values are much higher than elsewhere. This stands in contrast with many cities on the West Coast, for example. Second, whether high values of the ratio are the result of low property values rather than high rents depends on these classes of city. Third, once income, risk, and neighborhood characters are controlled for, tract-level racial makeup often has the "opposite" effect than what might be expected. High proportions of White residents, for example, have lower \textit{RPV} ratios, all else equal. All of these findings require more detailed investigation, however.

\section{Conclusion}

This study examines the phenomenon, common to many U.S. central cities, where rents do not fall proportionally to match low property values. The resulting gap means that renters in these areas pay relatively more for housing; these residents often have low incomes and are not White. The ratio of rents to property values has been referred to as "exploitation" elsewhare in the literature, but may be due to risk--since high vacancies or non-payment of rent incurs costs to landlords that is passed on to the renter.

Here, this \textit{RPV} ratio is calculated for 30 large U.S. cities and their surrounding MSAs, with a focus on the highest, "extreme" values. Mapping a subset of four core cities shows that this ratio is highest in mainly Black tracts in Milwaukee, Chicago, and Atlanta; San Francisco represents a different type of city with its own specific features.

Across cities, there are clear relationships between median \textit{RPV} ratios and a set of socioeconomic variables; while this ratio is positively associated with the rental vacancy rate, cities such as Buffalo, Cleveland, and Detroit stand out as "outliers." Correlations with the median percentages of White or Black residents are weaker at the MSA level, however.

Ranking each core city by its mean \textit{RPV} ratio shows the highest median and 95\% quantiles for Detroit and other "Rust Belt" cities, with many higher income cities--as well as New York and L.A.--clustered near the bottom. This is likely due to differences in property values in each type of city. This study does isolate exactly where renters are most likely to pay the highest rents relative to property values, however. One interesting finding is that at the MSA level, some Southern cities have high median \textit{RPV} values--which is not shown to be the case for their core cities.

A spatial regression model is then used to examine variation in the z-scores of the \textit{RPV} ratio for each city. Including median income, median propery values, the overall vacancy rate, and the rental vacancy rate, these first two variables have a great deal of explanatory power in the model. The impact of vacancies is more mixed. But the percentages of White or Black residents in each tract--which have significant coefficients in about half of the core cities--tend not to support the hypothesis that racial exploitation is present. But further analysis is necessary to make any conclusive claims, particularly involving the role of rent assistance.

These findings might be useful for both local governments and community leaders who wish to stem the influx of property owners who buy rental properties in central-city neighborhoods in the hope of profiting from high rent-to-property-value ratios. By identifying specific neighborhoods, incentives could be targeted for local investors or owner-occupants to purchase and renovate the housing stock. At a national level, types of cities could be further identified, so that housing policy in "Rust Belt" and similar cities could improve residents' lives.

\begin{figure}[b]
\hfill

\caption{Milwaukee Tract Values.}
\includegraphics[width=.2\textwidth]{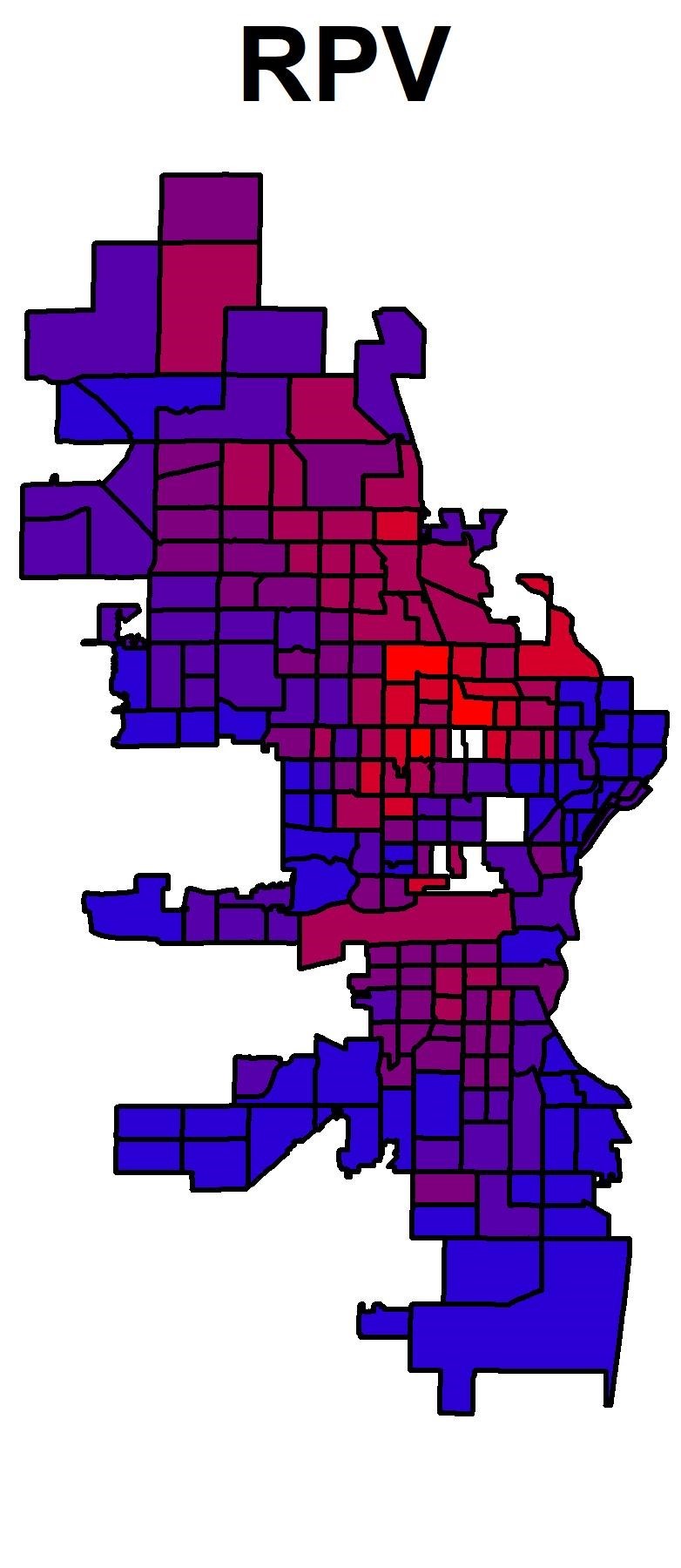}
\includegraphics[width=.2\textwidth]{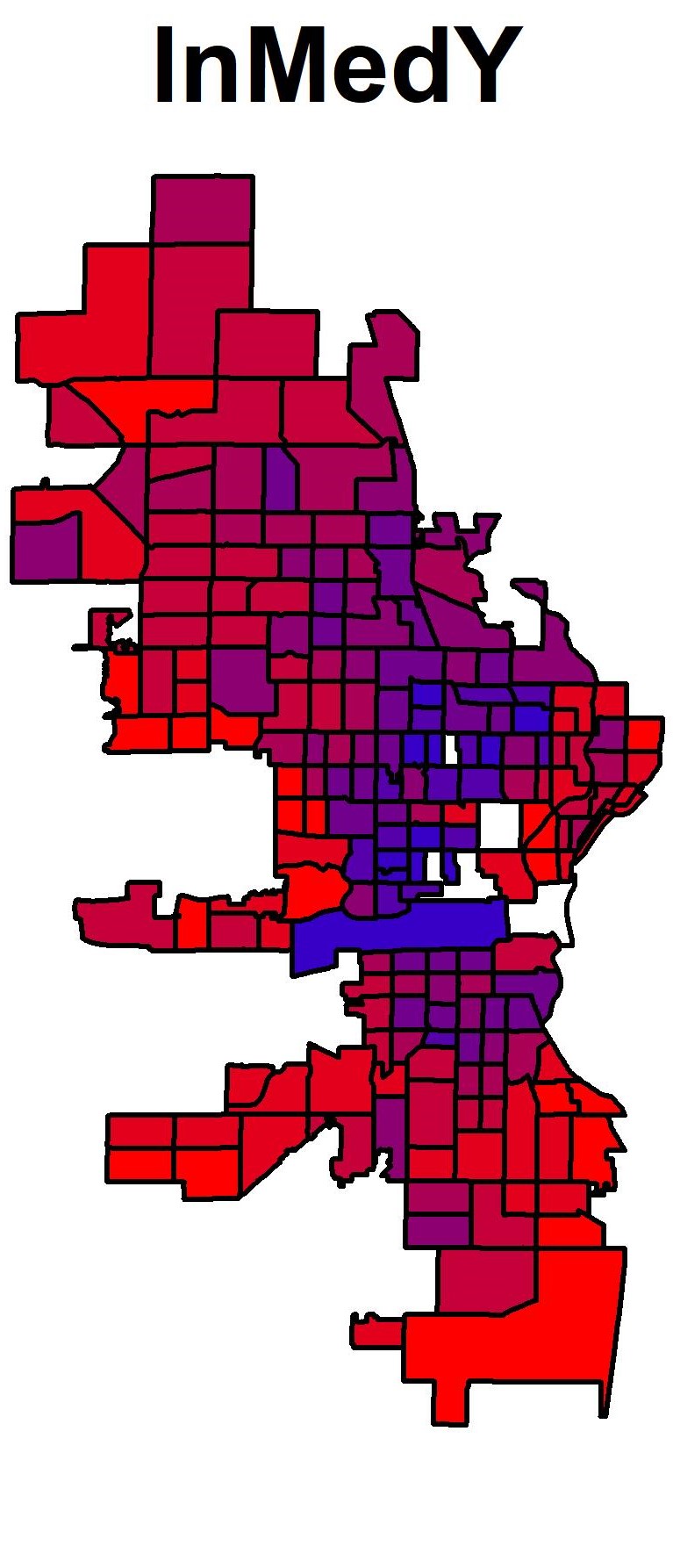}
\includegraphics[width=.2\textwidth]{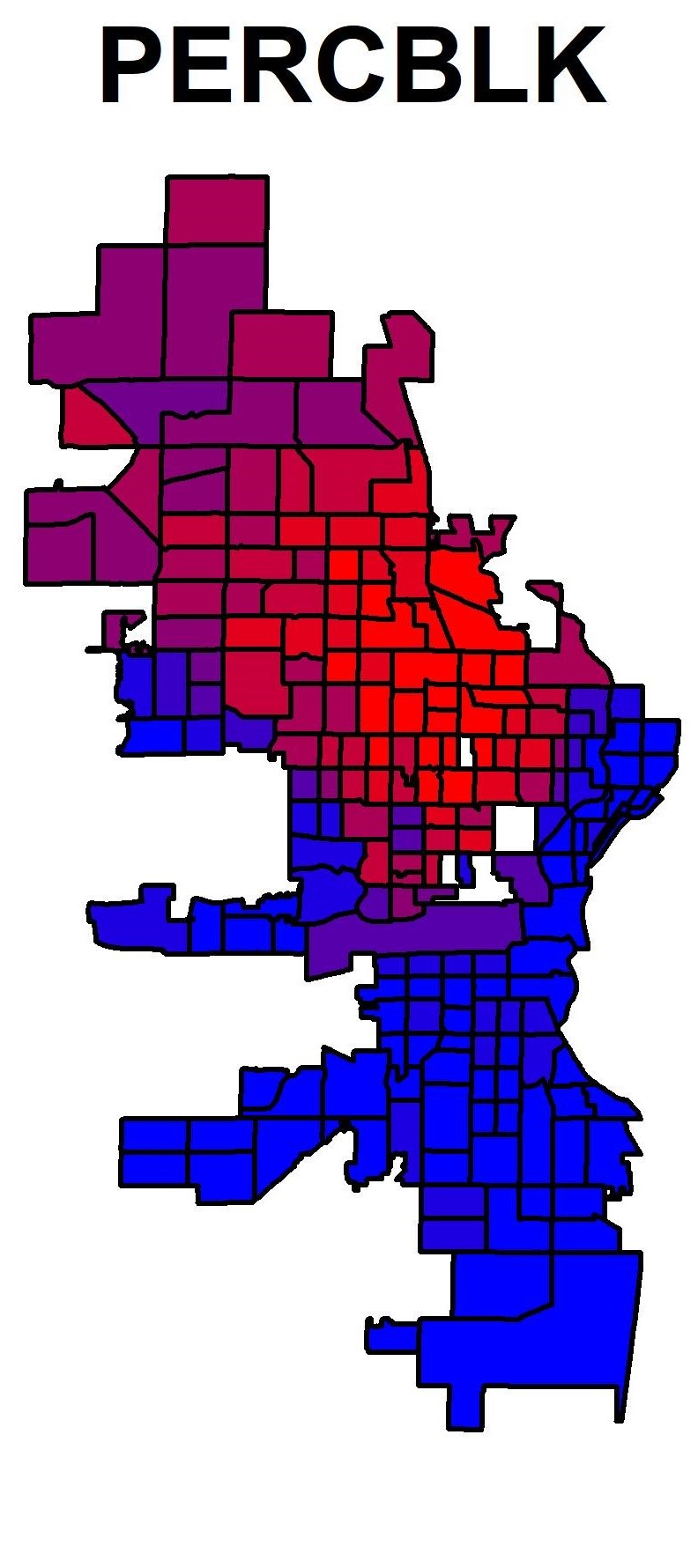}
\includegraphics[width=.2\textwidth]{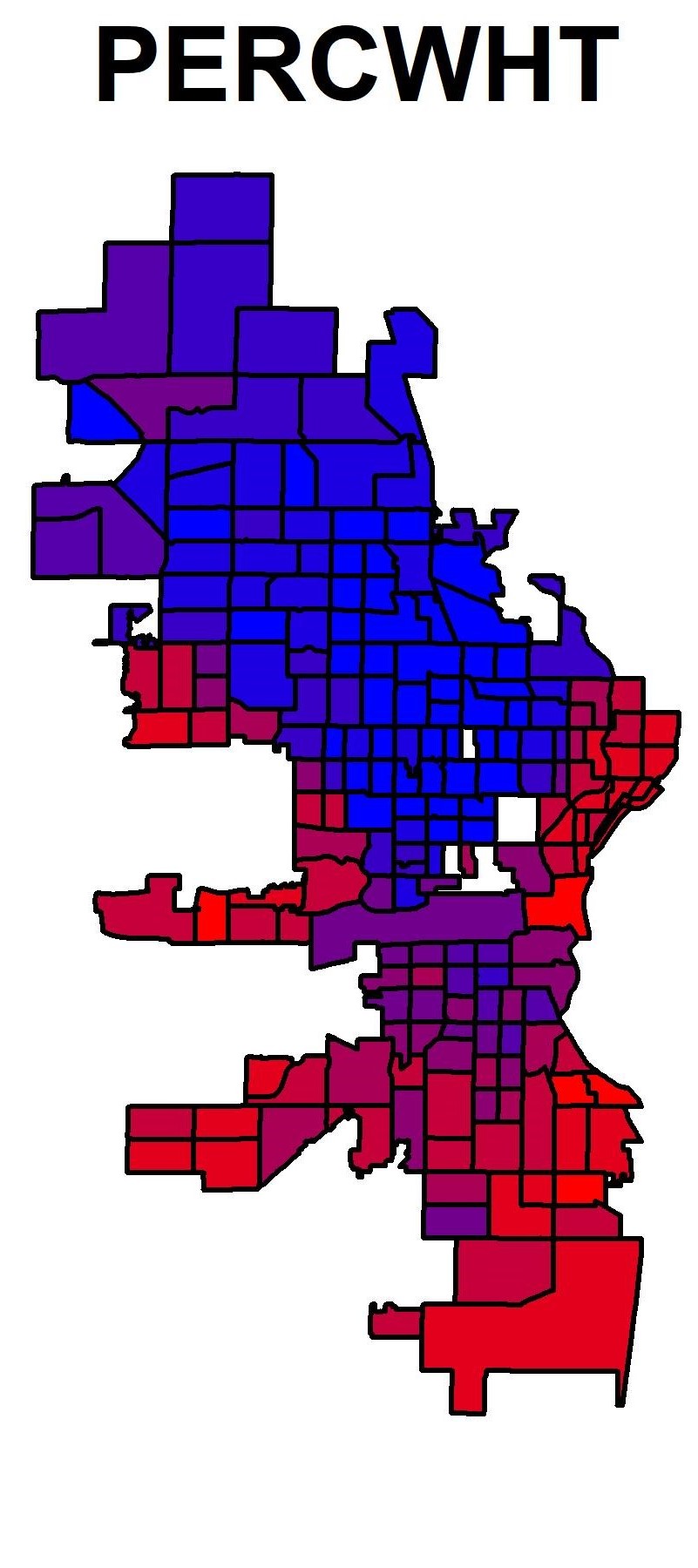}
\includegraphics[width=1\textwidth]{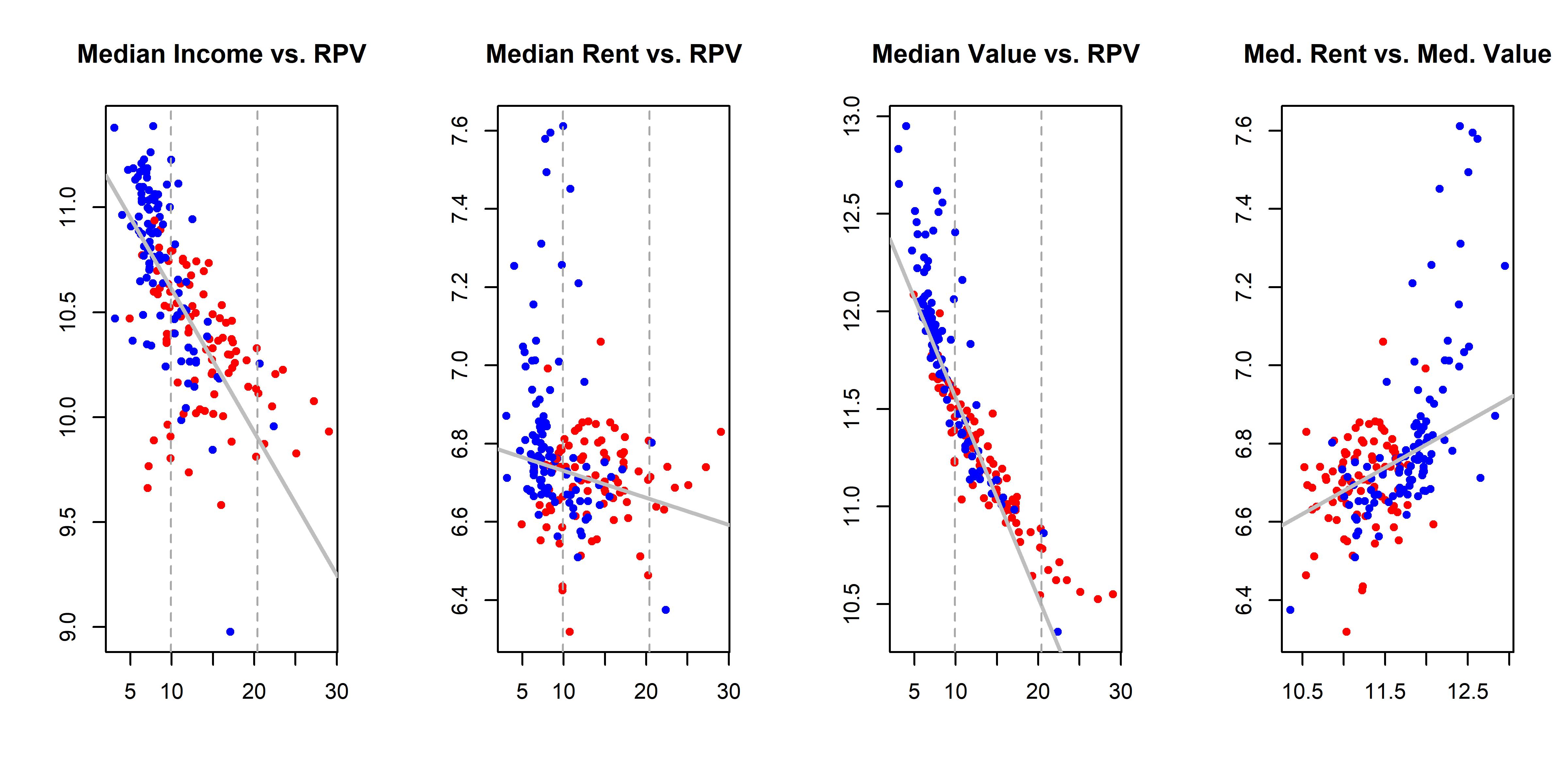}
\caption*{Top panel: Blue = low values; red = higher.\\
Median values in natural logarithms.\\
Bottom panel: Red dots: $\geq$ 50\% Black residents. Vertical lines = median and 95\% RPV quantile values.\\
Regression lines based on the Thiel-Sen (1968) nonparametric estimator.
}
\end{figure}

\begin{figure}[b]
\hfill
\caption{Chicago Tract Values.}
\includegraphics[width=.2\textwidth]{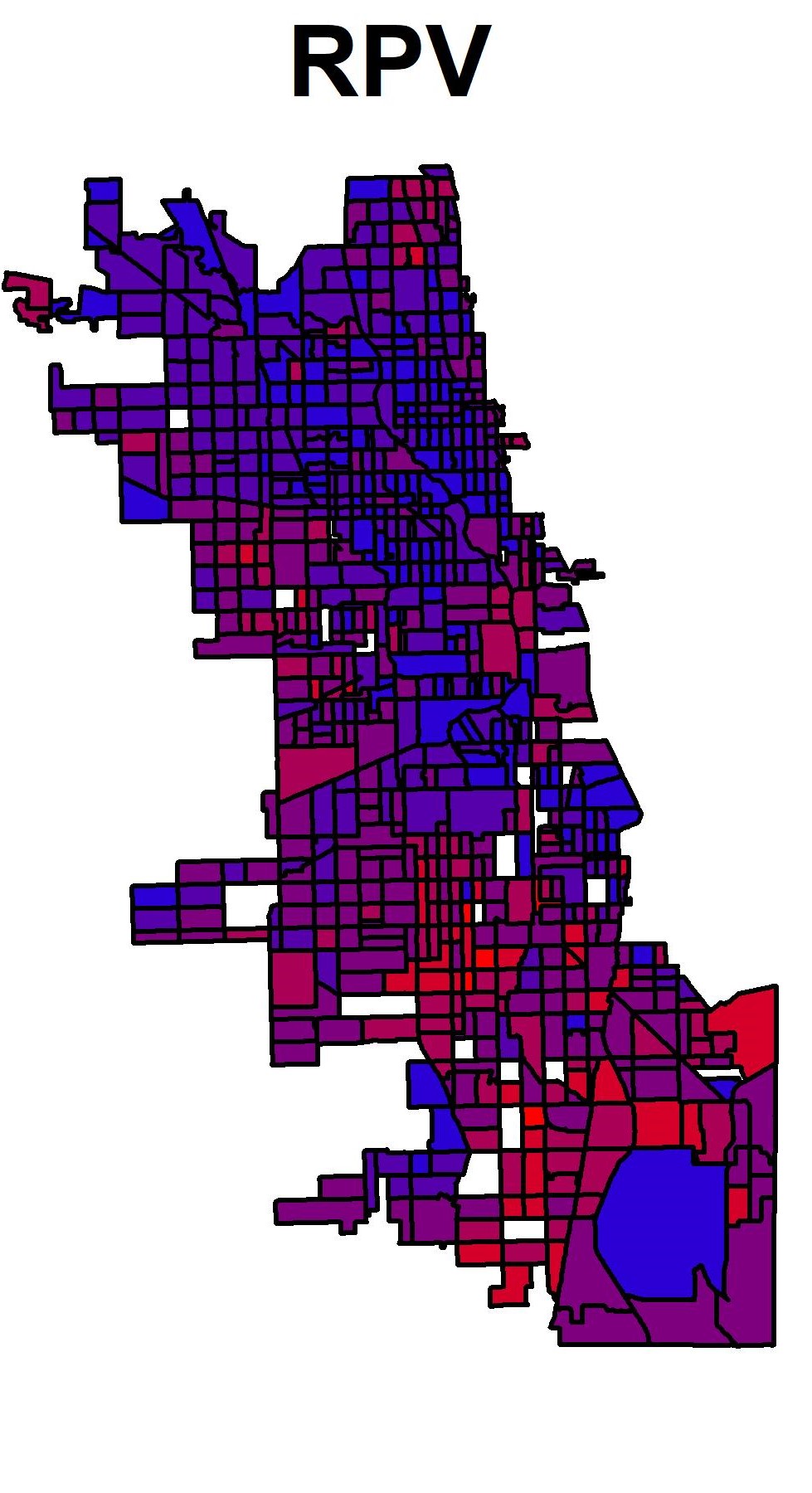}
\includegraphics[width=.2\textwidth]{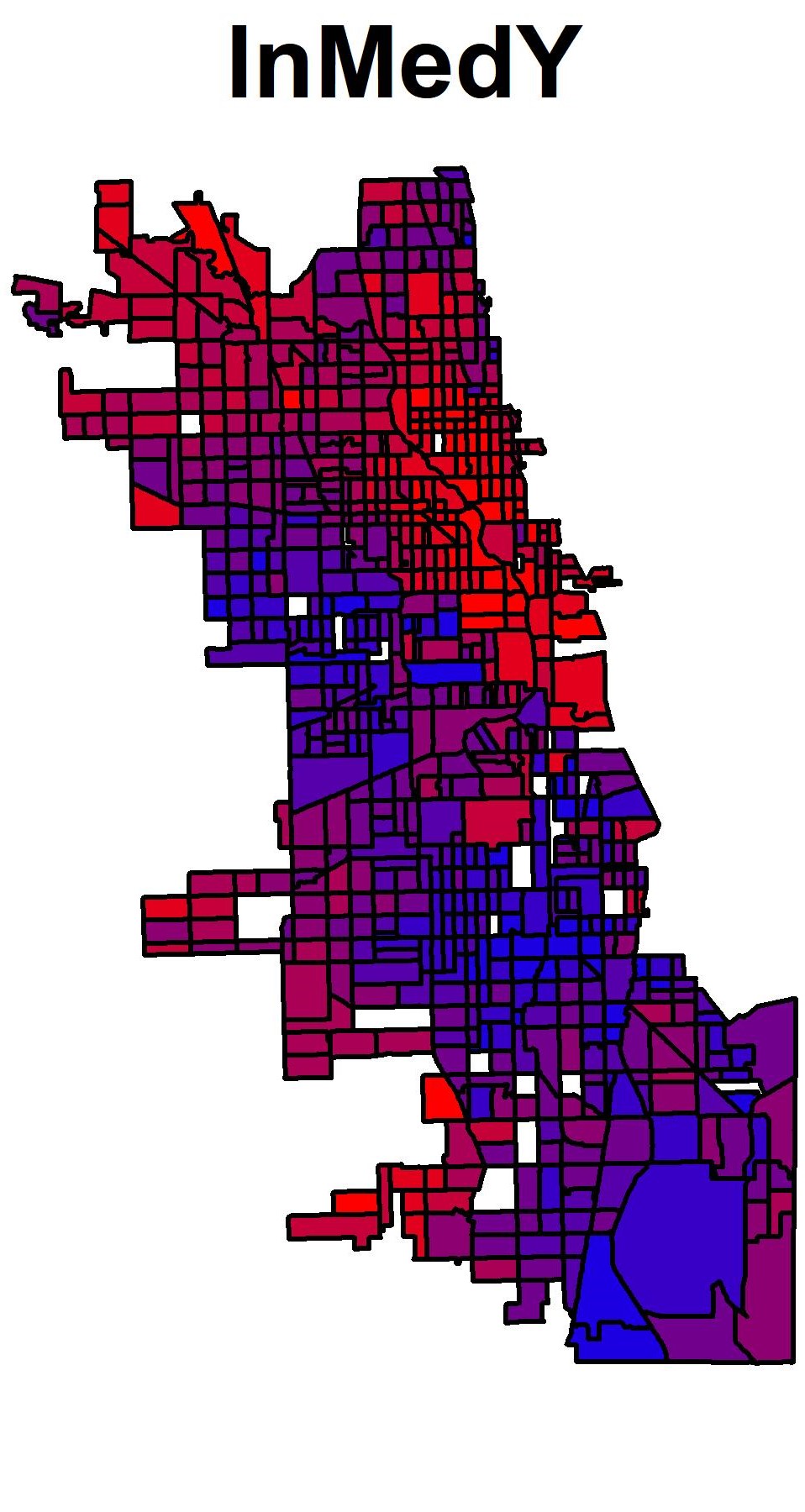}
\includegraphics[width=.2\textwidth]{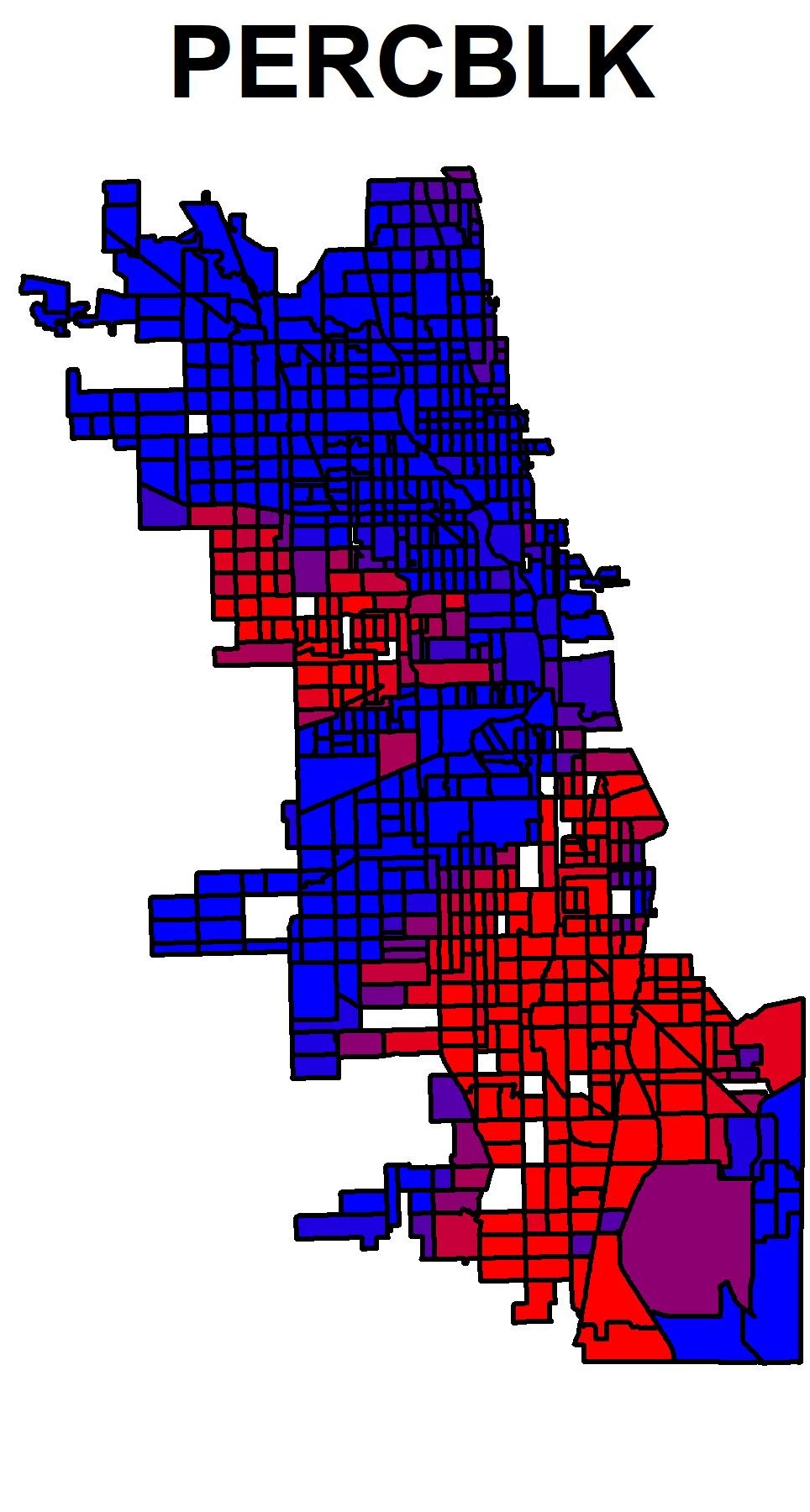}
\includegraphics[width=.2\textwidth]{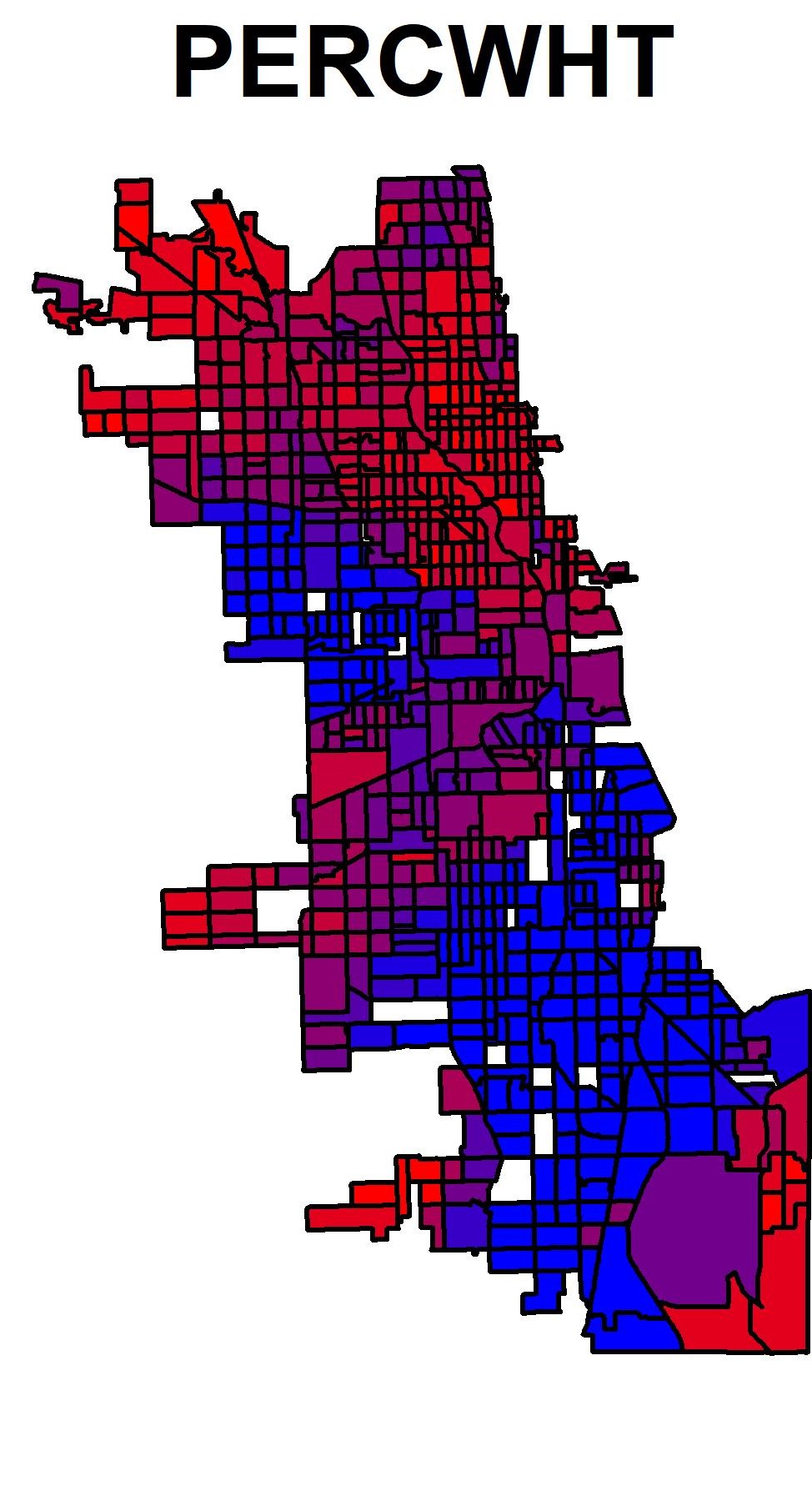}
\includegraphics[width=1\textwidth]{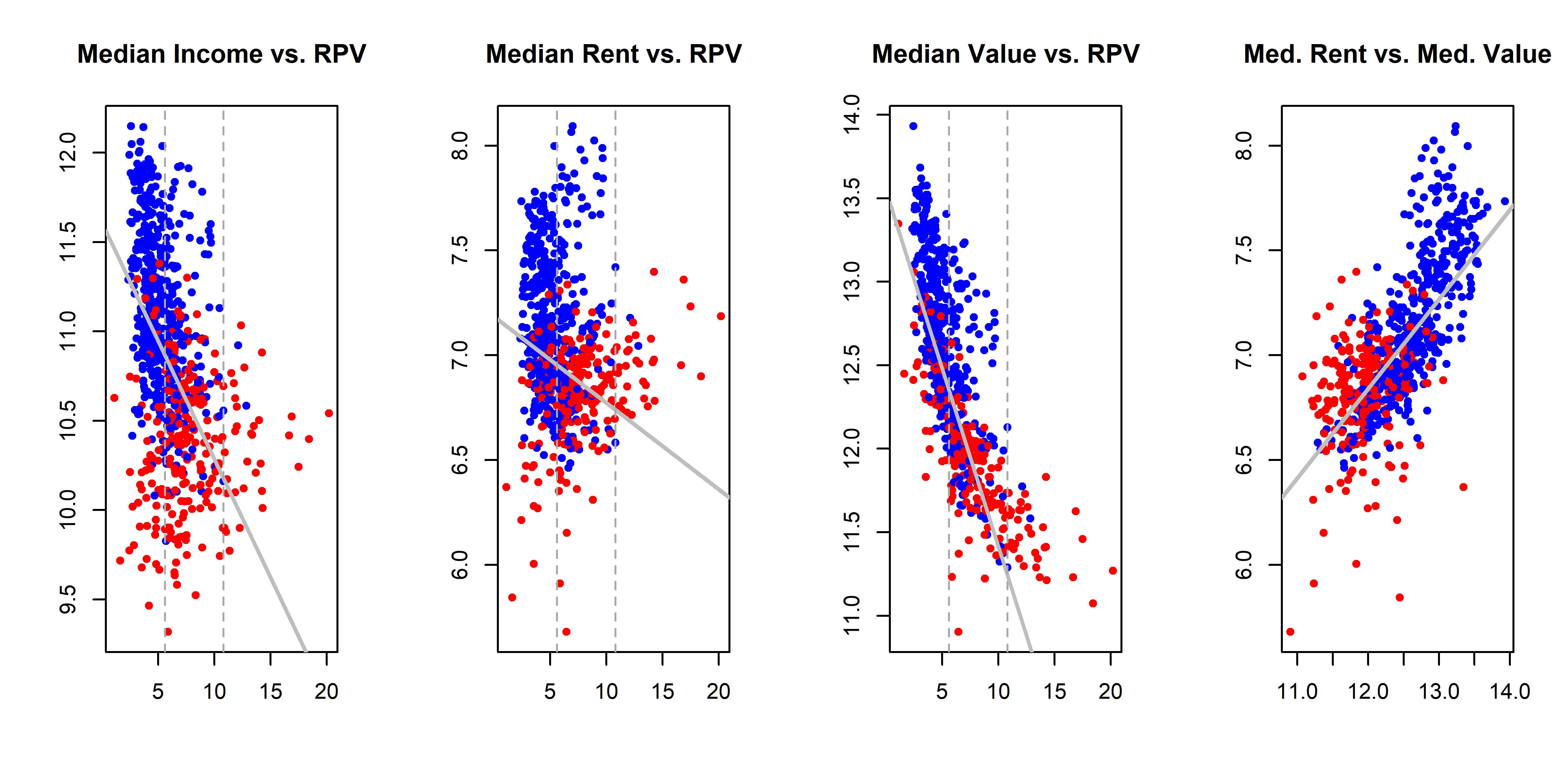}
\caption*{Top panel: Blue = low values; red = higher.\\
Median values in natural logarithms.\\
Bottom panel: Red dots: $\geq$ 50\% Black residents. Vertical lines = median and 95\% RPV quantile values.\\
Regression lines based on the Thiel-Sen (1968) nonparametric estimator.
}
\end{figure}

\begin{figure}[b]
\hfill
\caption{Atlanta Tract Values.}
\includegraphics[width=.2\textwidth]{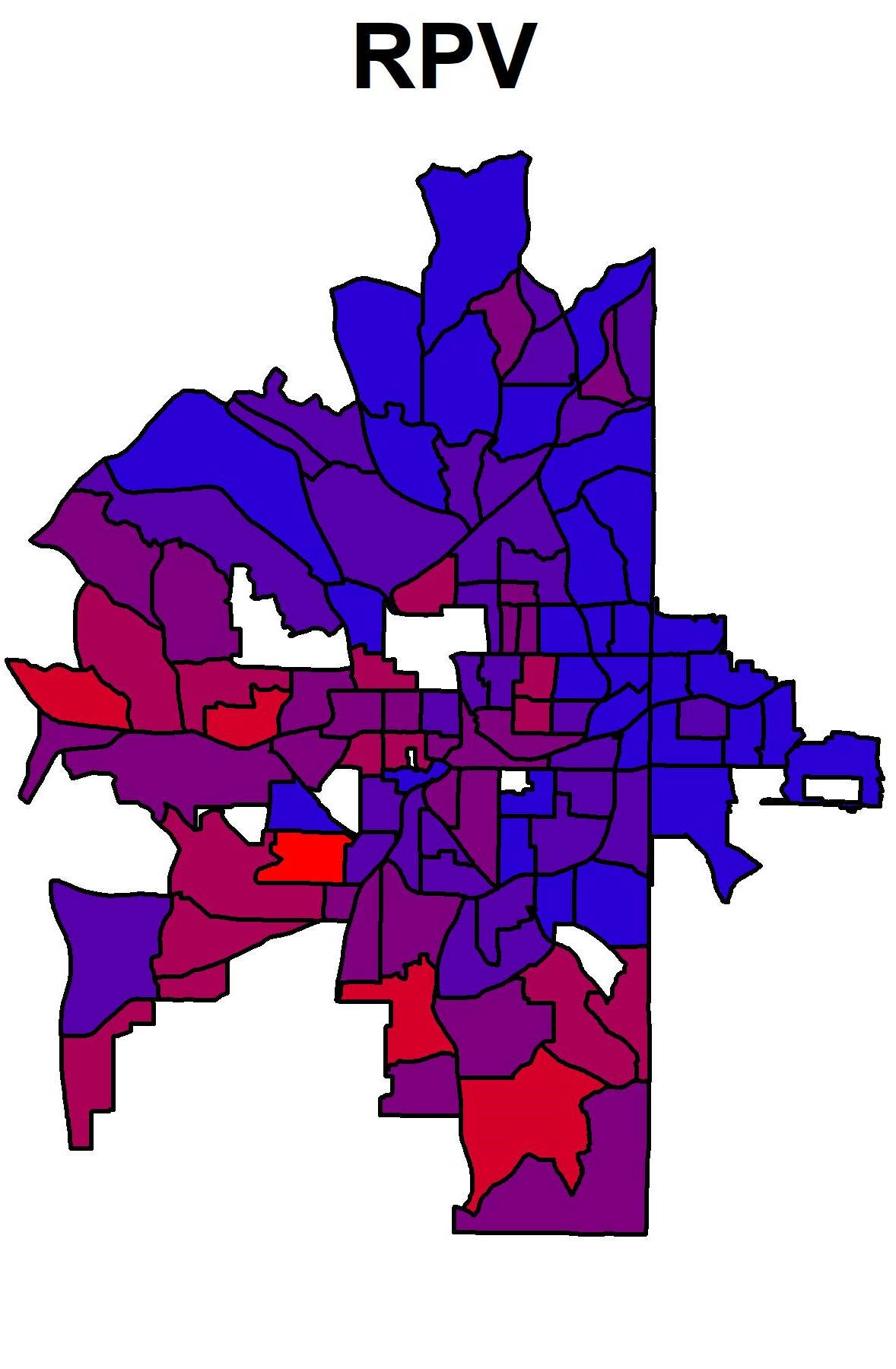}
\includegraphics[width=.2\textwidth]{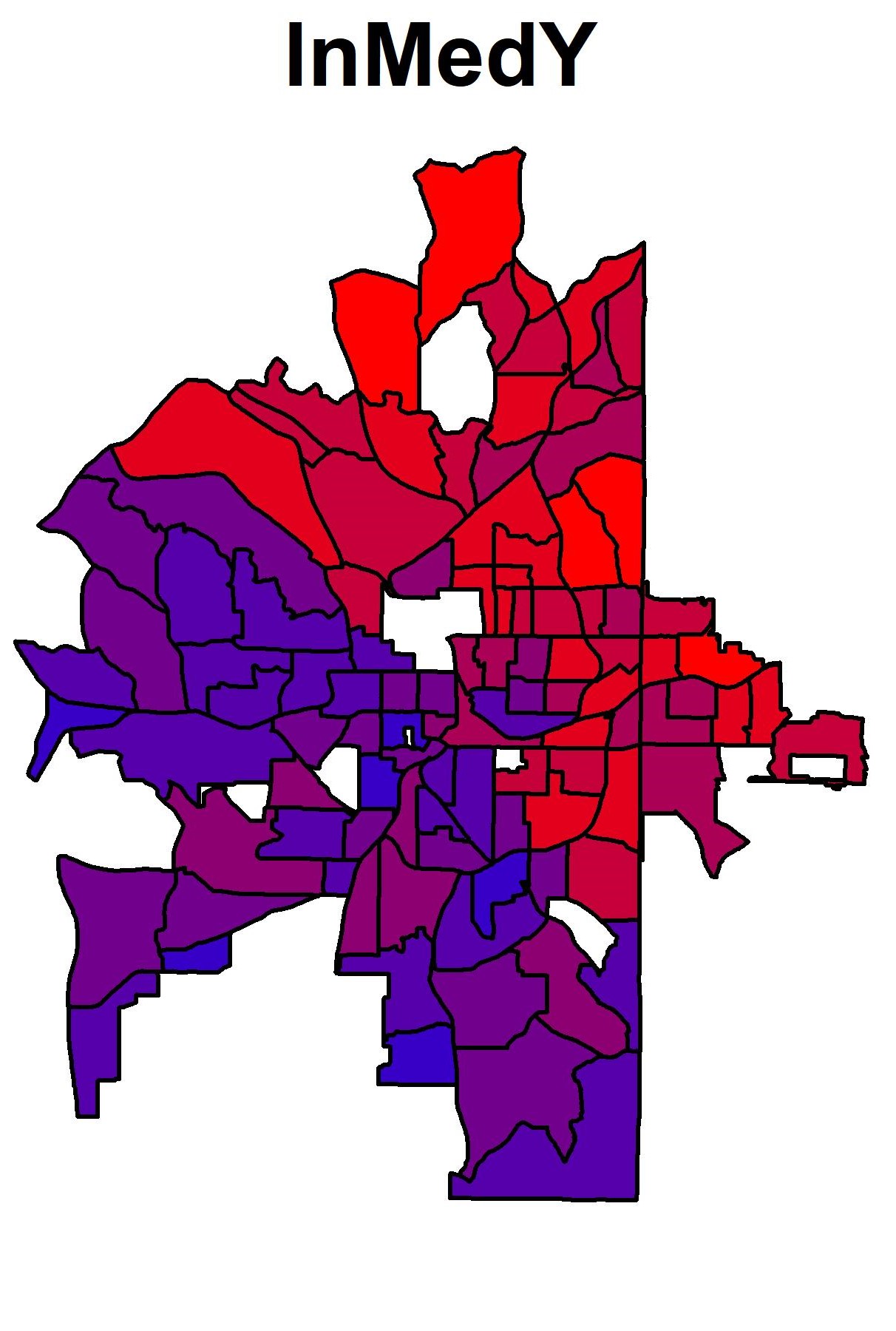}
\includegraphics[width=.2\textwidth]{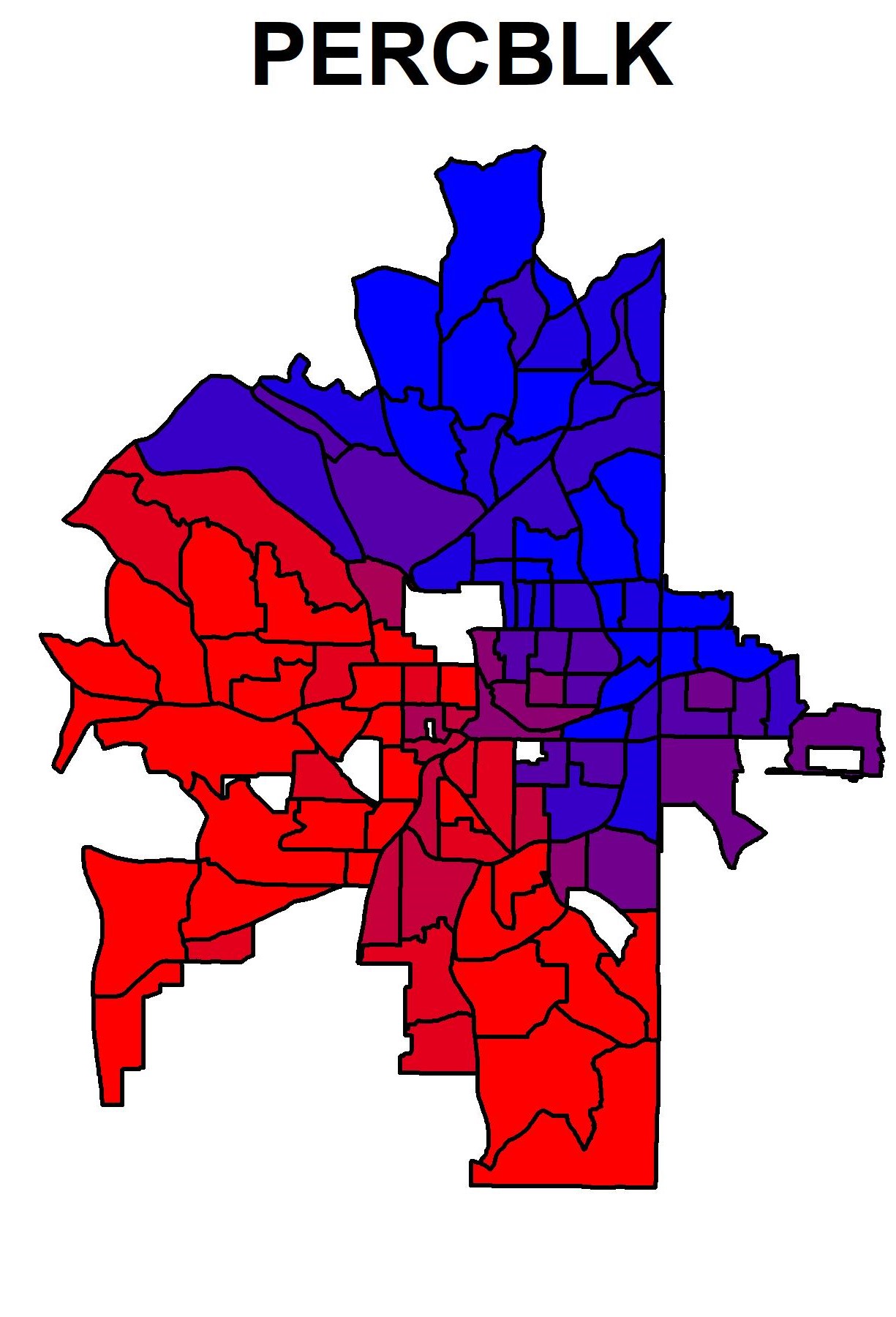}
\includegraphics[width=.2\textwidth]{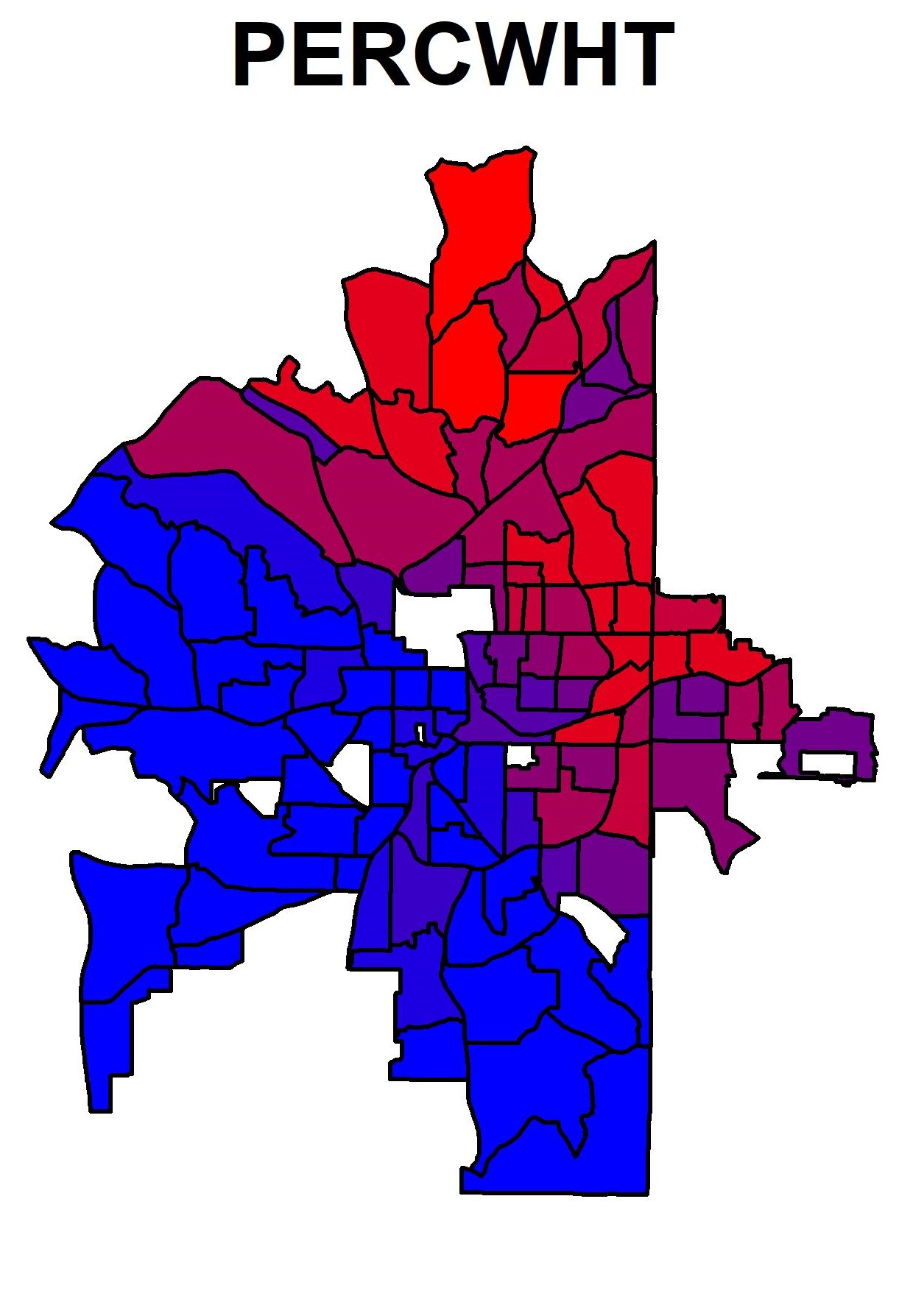}
\includegraphics[width=1\textwidth]{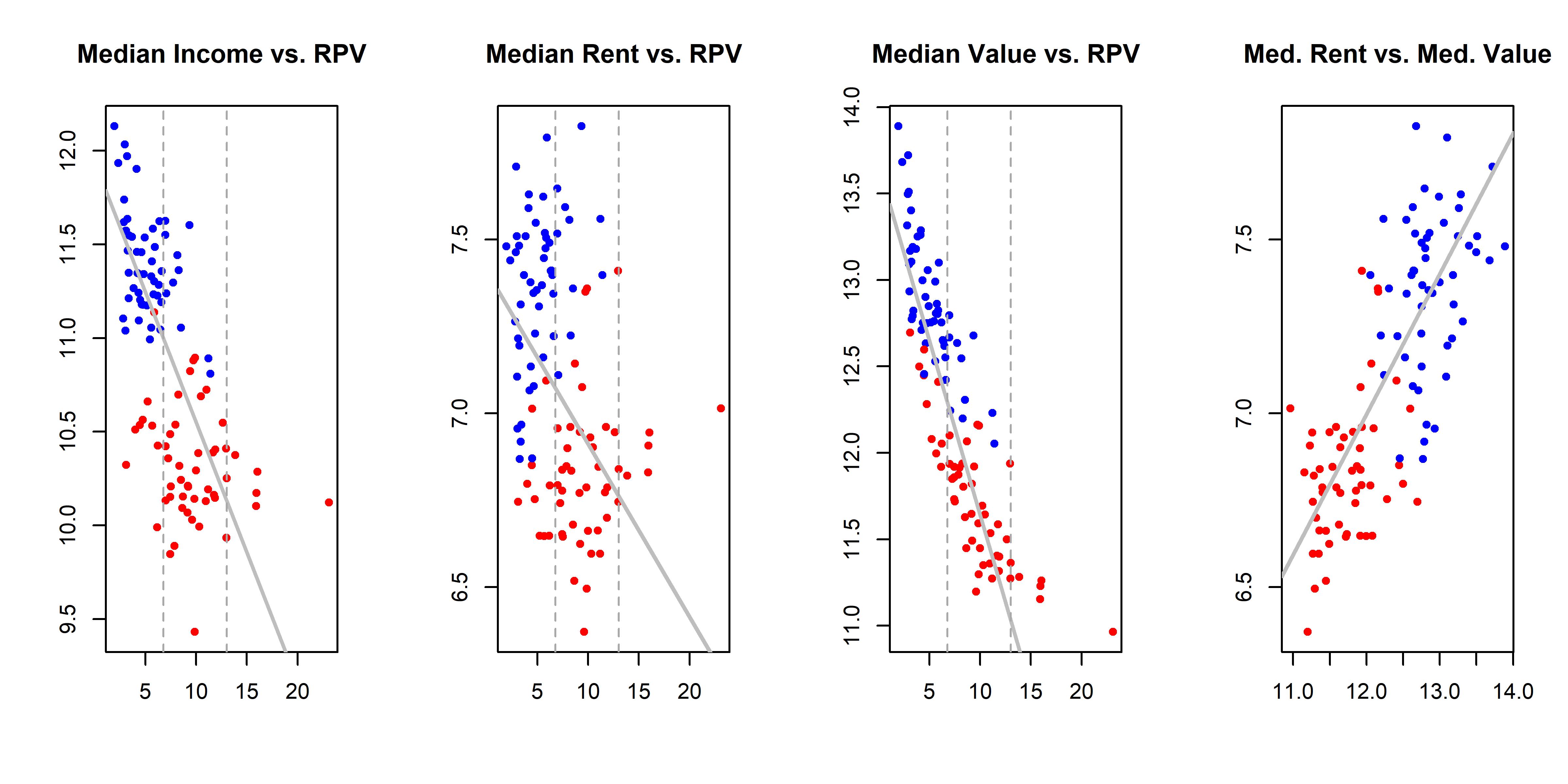}
\caption*{Top panel: Blue = low values; red = higher.\\
Median values in natural logarithms.\\
Bottom panel: Red dots: $\geq$ 50\% Black residents. Vertical lines = median and 95\% RPV quantile values.\\
Regression lines based on the Thiel-Sen (1968) nonparametric estimator.
}
\end{figure}

\begin{figure}[b]
\hfill
\caption{San Francisco Tract Values.}
\includegraphics[width=.2\textwidth]{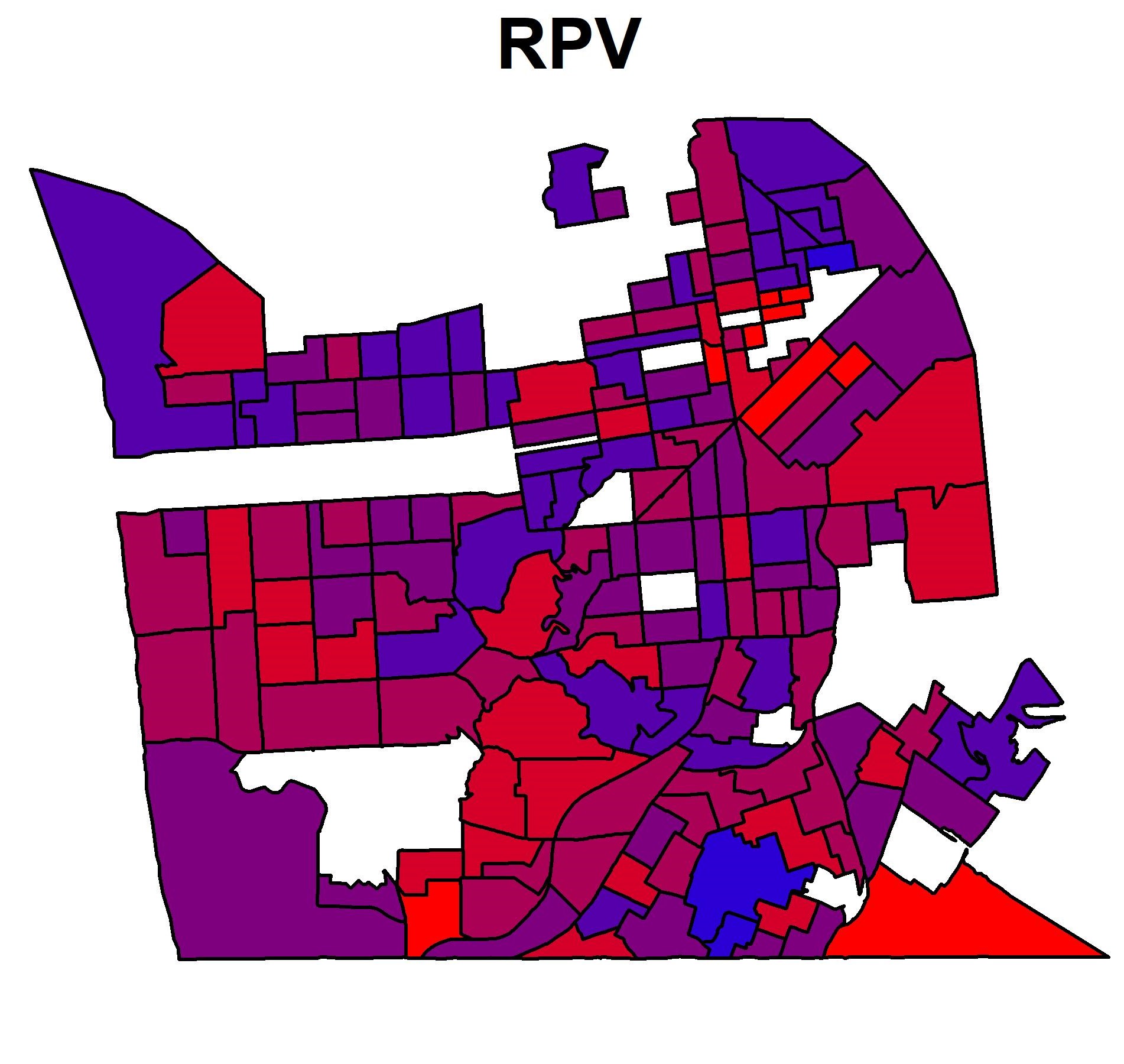}
\includegraphics[width=.2\textwidth]{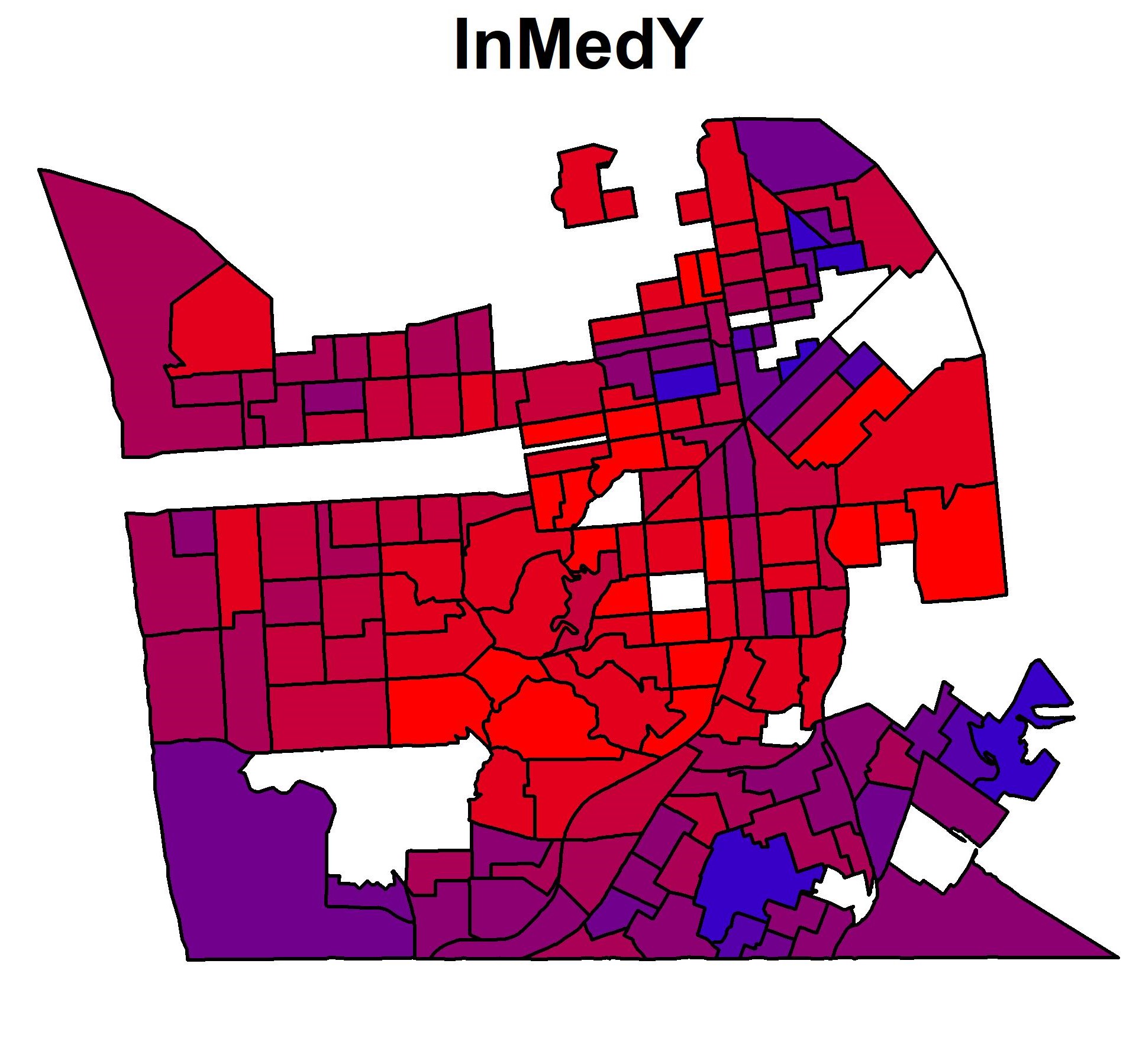}
\includegraphics[width=.2\textwidth]{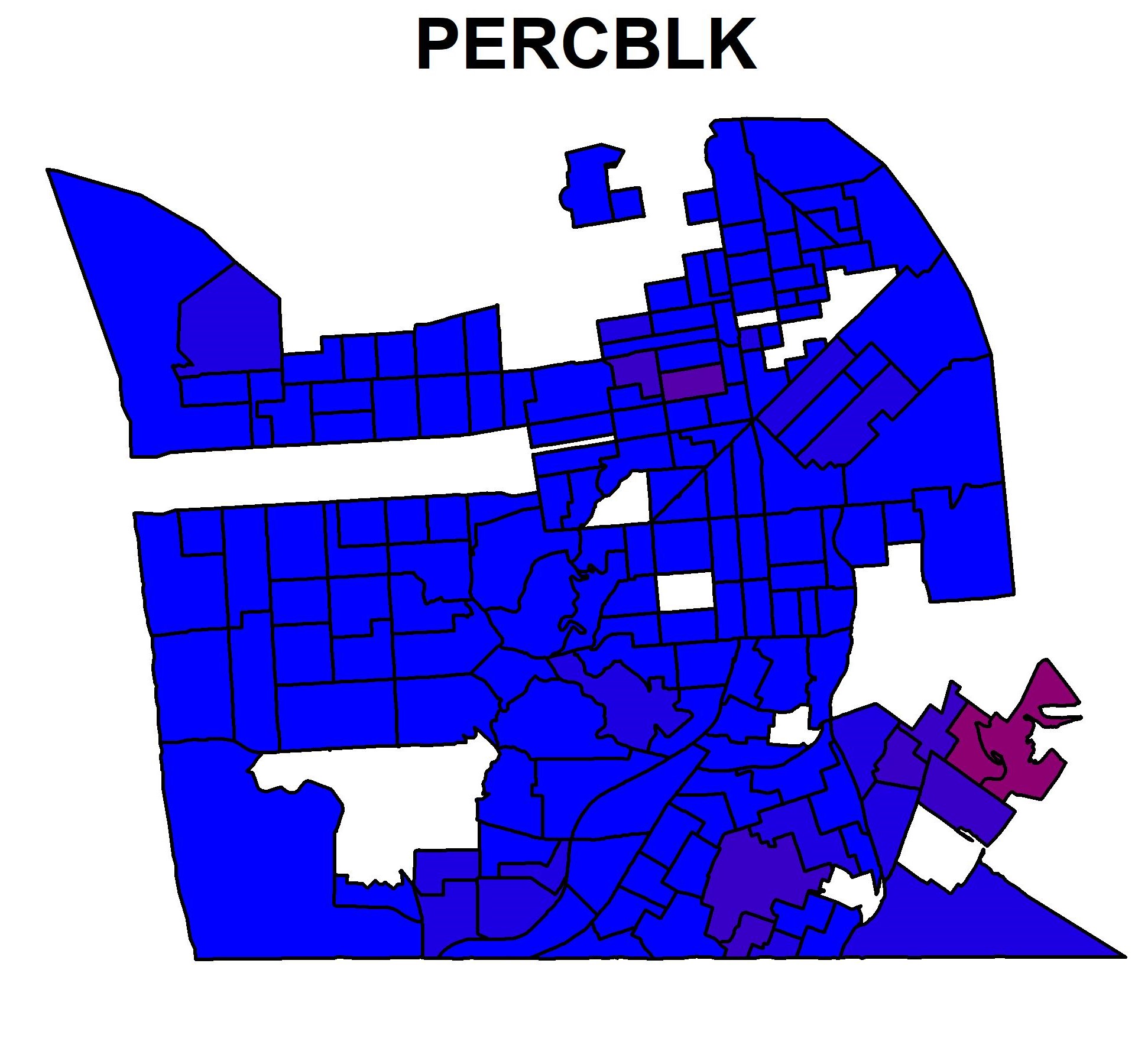}
\includegraphics[width=.2\textwidth]{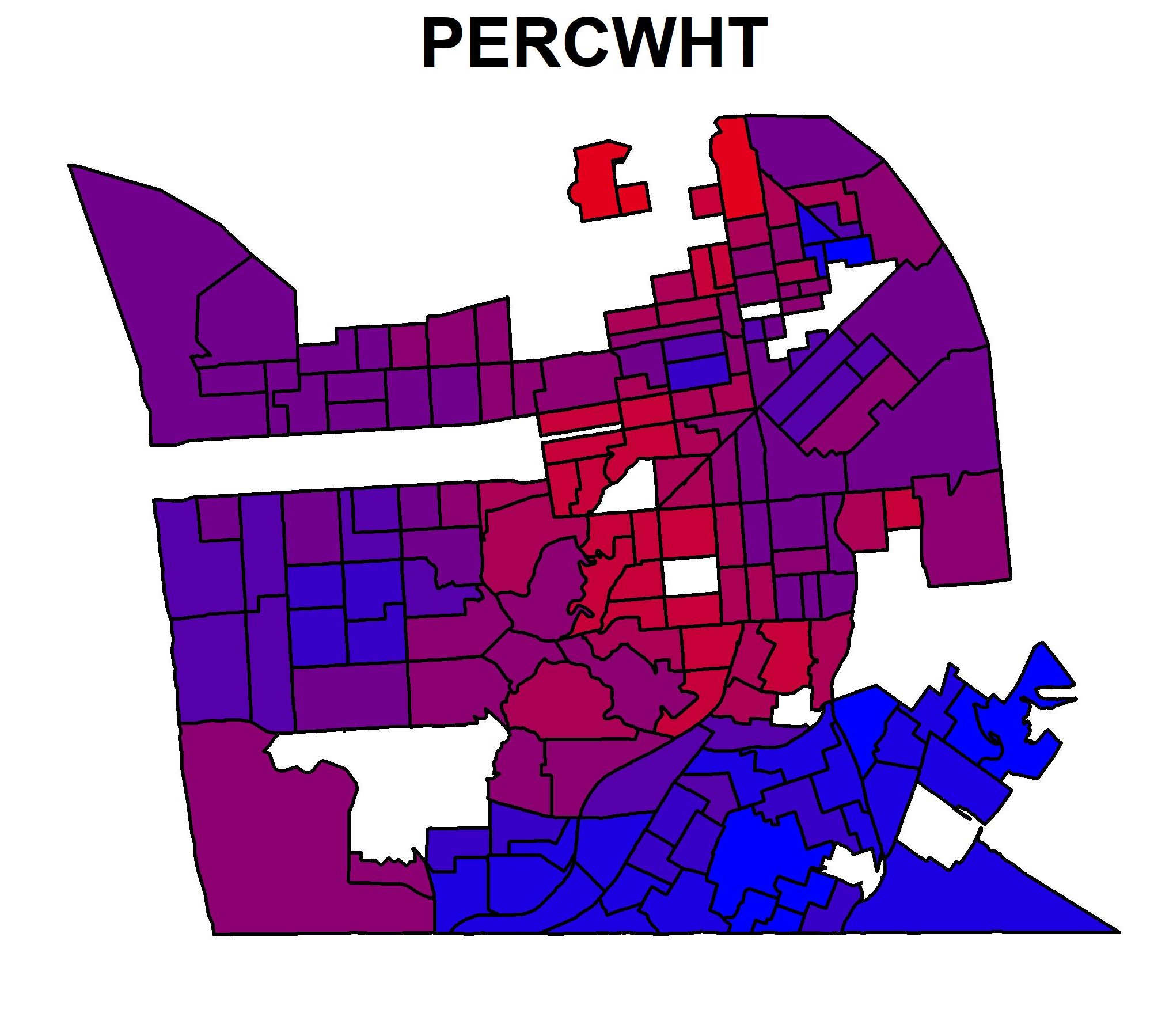}
\includegraphics[width=1\textwidth]{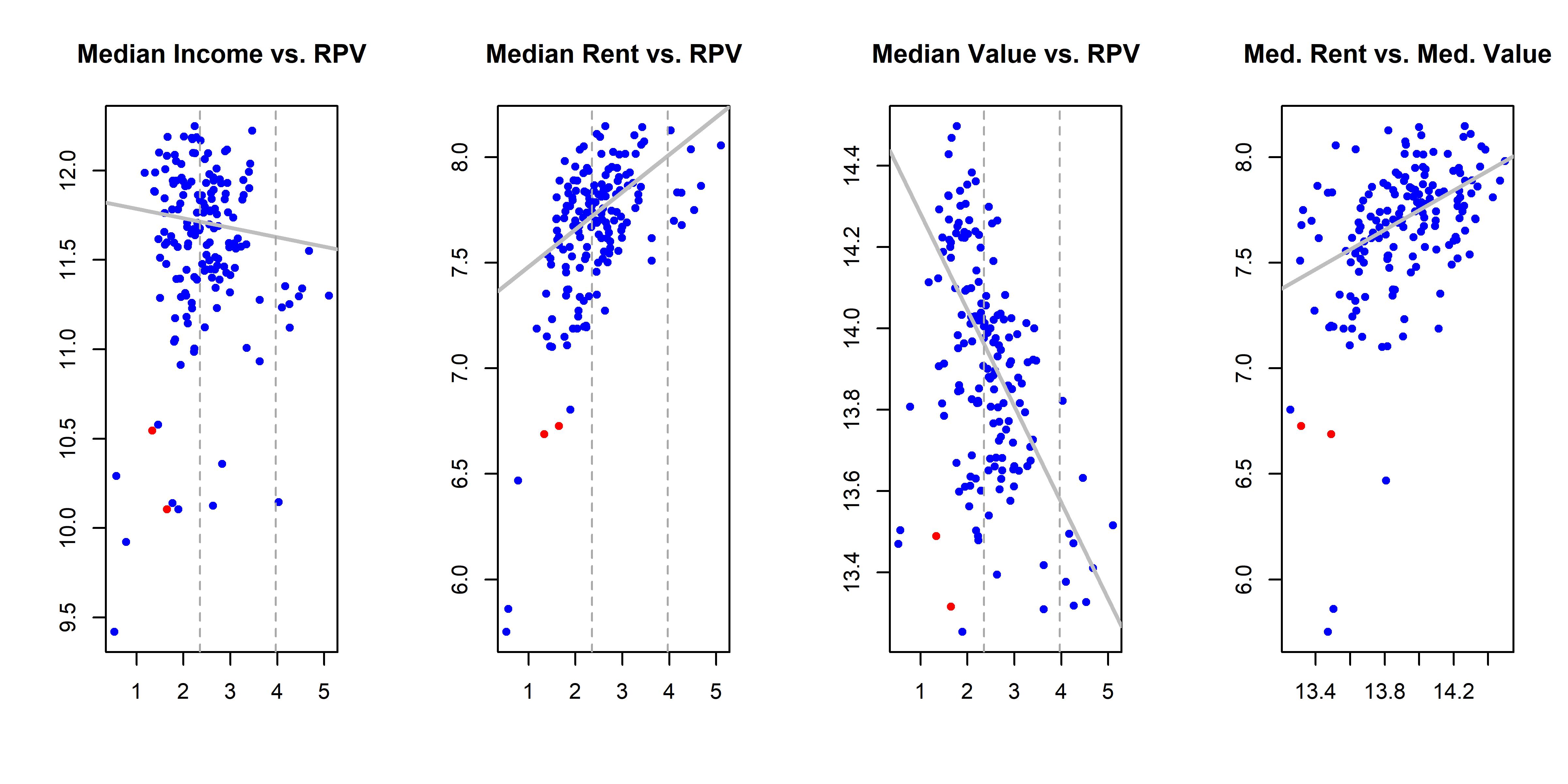}
\caption*{Top panel: Blue = low values; red = higher.\\
Median values in natural logarithms.\\
Bottom panel: Red dots: $\geq$ 50\% Black residents. Vertical lines = median and 95\% RPV quantile values.\\
Regression lines based on the Thiel-Sen (1968) nonparametric estimator.
}
\end{figure}

\begin{figure}[b]
    \begin{adjustwidth}{-.75in}{-.5in}  
        \begin{center}
\hfill
\caption{Median RPV (x-axis) versus Other Variables, 30 Core Cities.}
\includegraphics[width=1.4\textwidth]{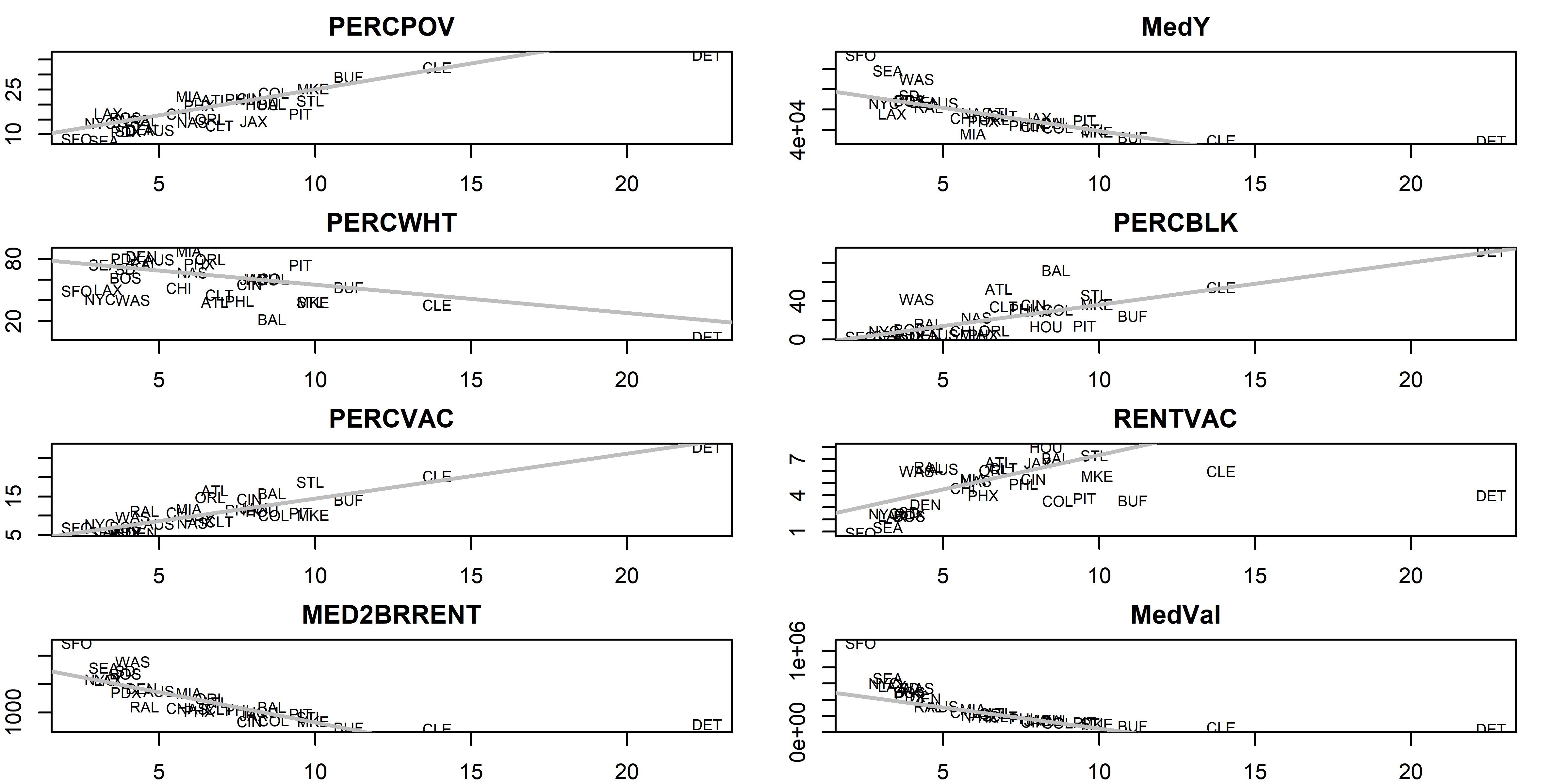}
\caption*{Regression lines based on the Thiel-Sen (1968) nonparametric estimator.}
        \end{center}
\end{adjustwidth}
\end{figure}

\begin{figure}[b]
    \begin{adjustwidth}{-.75in}{-.5in}  
\hfill
\caption{Median RPV versus Other Variables, 30 MSAs.}
\includegraphics[width=1.4\textwidth]{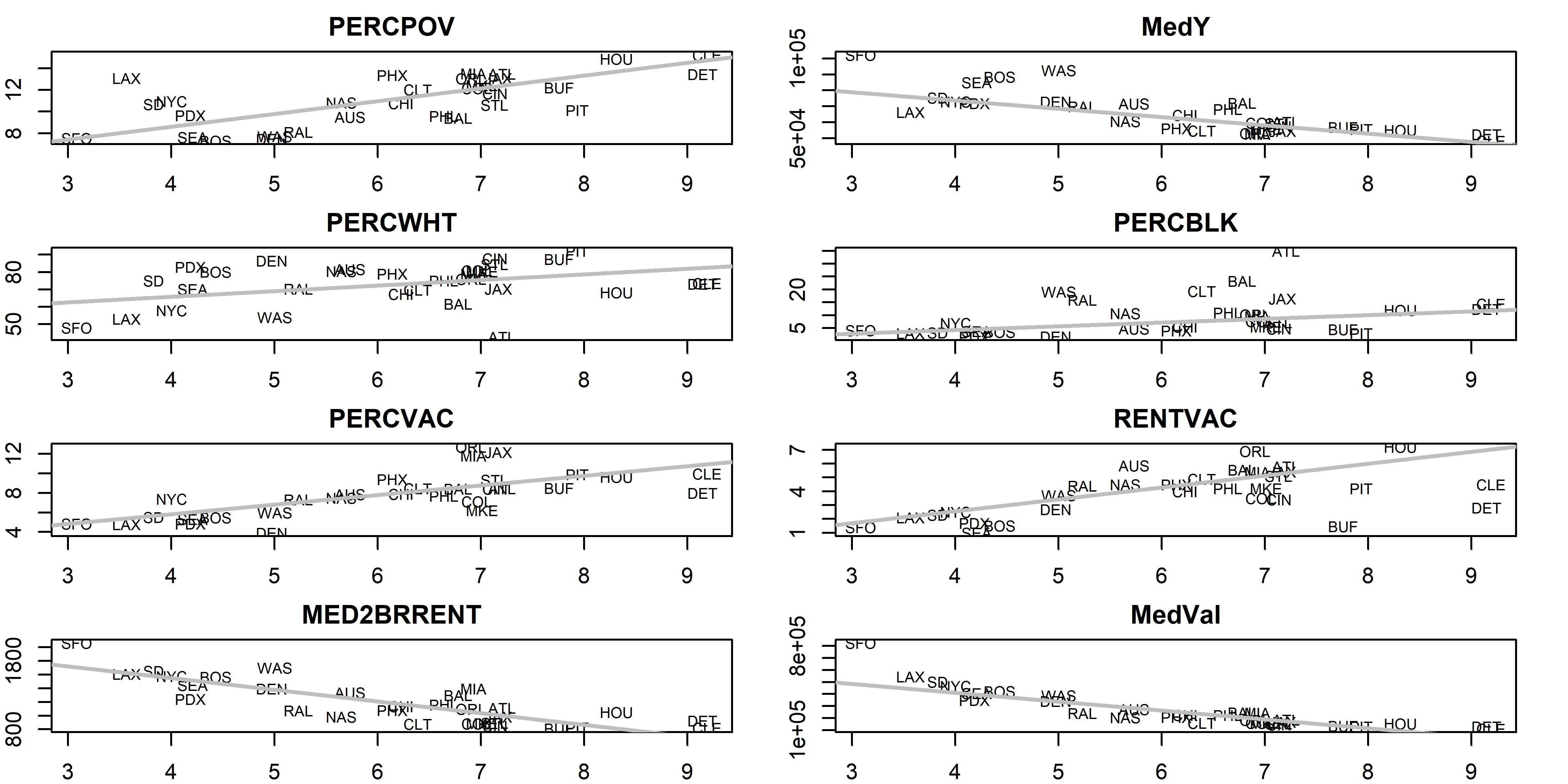}
\caption*{
Regression lines based on the Thiel-Sen (1968) nonparametric estimator.
}
\end{adjustwidth}
\end{figure}

\begin{figure}[b]
\hfill
\caption{RPV Quantiles for 30 Core Cities.}
\includegraphics[width=1\textwidth]{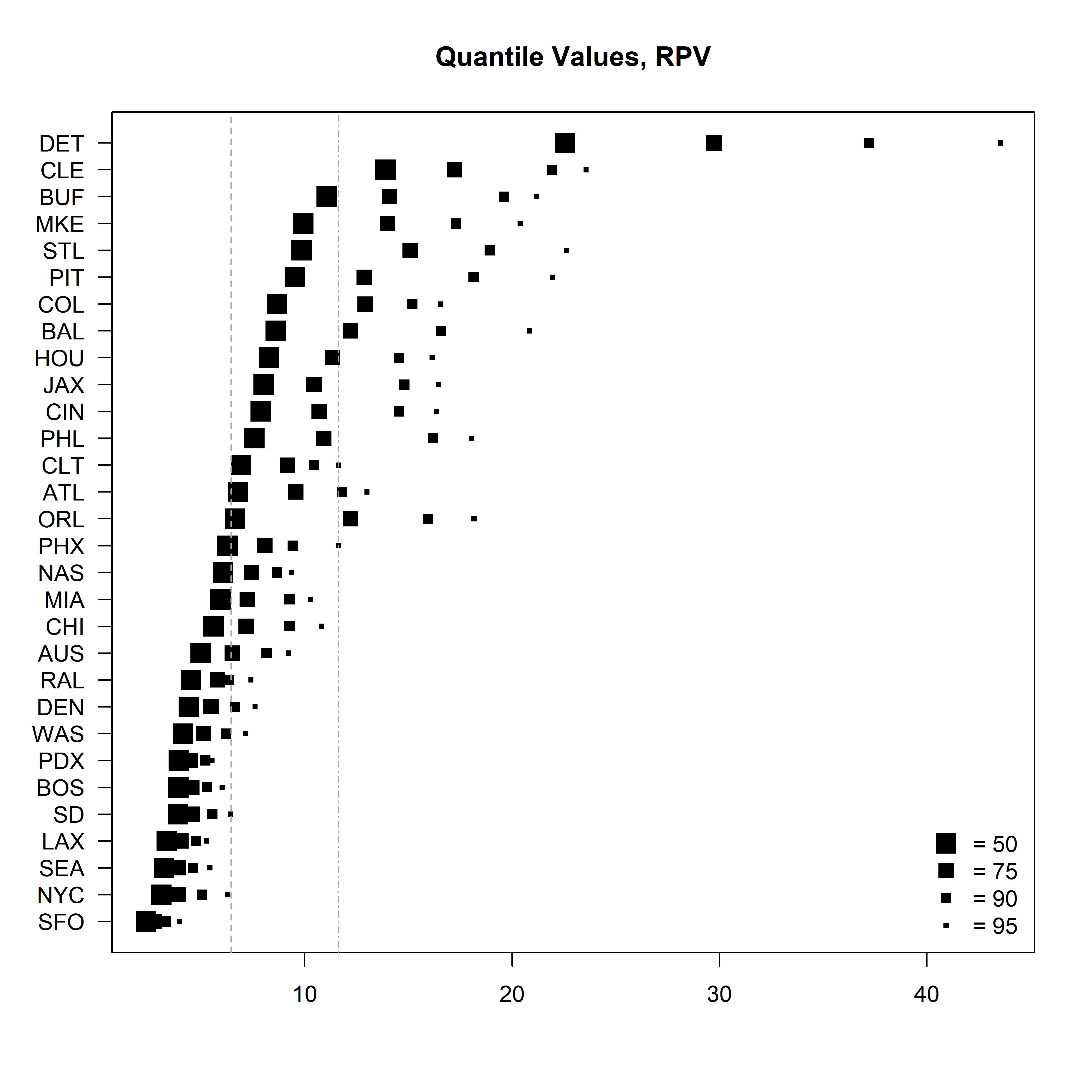}
\caption*{Vertical lines = Sample-wide Medians for the 50\% and 95\% quantiles.}
\end{figure}

\begin{figure}[b]
\hfill
\caption{RPV Quantiles for 30 Core MSAs.}
\includegraphics[width=1\textwidth]{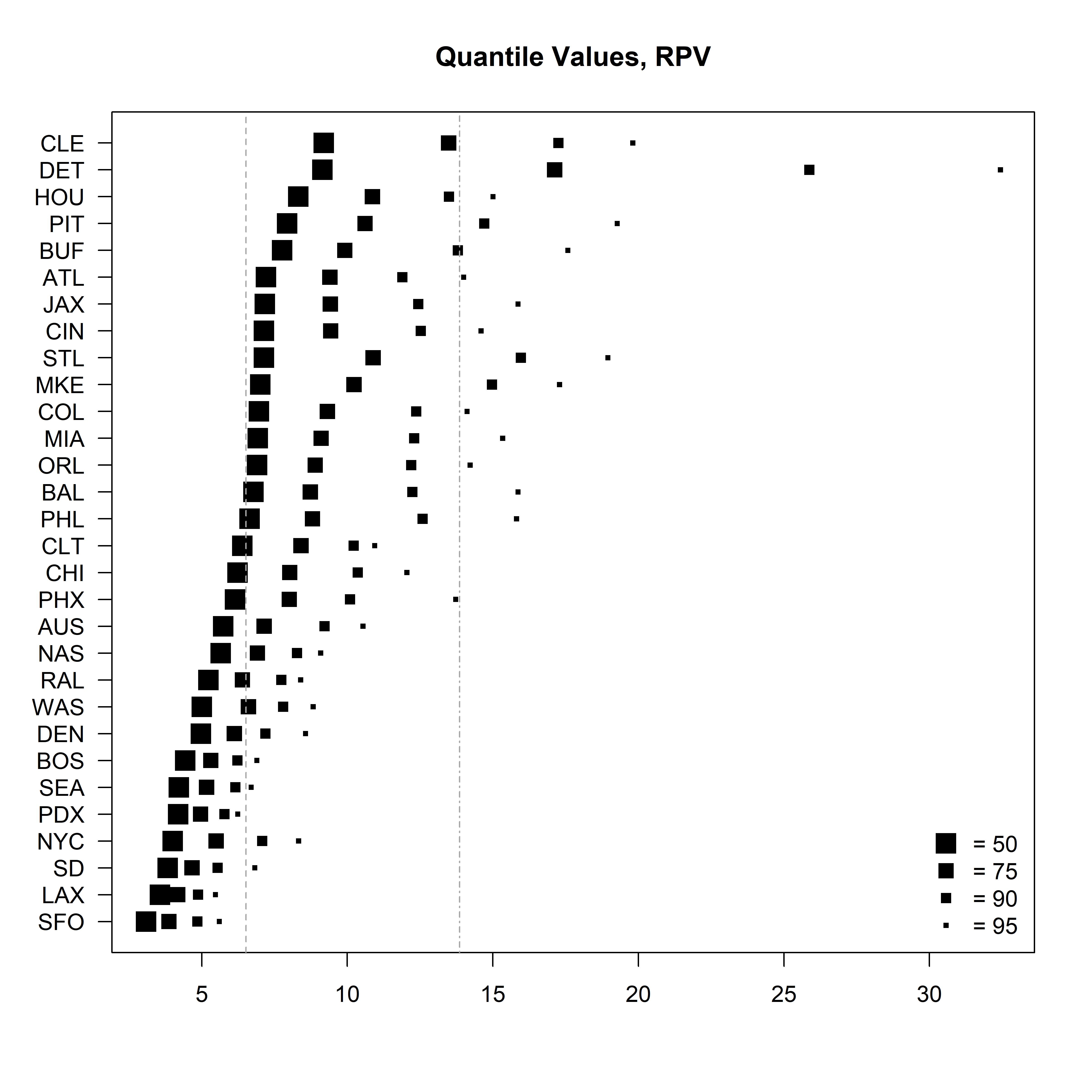}
\caption*{Vertical lines = Sample-wide Medians for the 50\% and 95\% quantiles.}
\end{figure}

\begin{table}[b]
    \begin{adjustwidth}{-.5in}{-.5in}  
\caption{Median Core-City Statistics for 30 MSAs.}
        \begin{center}

\begin{tabular}{lrrrrrrrrr}

CITY&N&\%POV&MedY&\%WHT&\%BLK&\%VAC&RENTVAC&MEDRENT&MedVal\\
\hline
ATL&106&21.6&56666&38.4&52.6&16.6&6.8&1177&233350\\
AUS&125&11.5&66296&79.4&4.9&7.9&6.2&1381&317700\\
BAL&172&20.1&46907&21.8&72.2&15.8&7.1&1095&145600\\
BOS&150&15.6&68764&61.7&10.6&7.1&2.3&1684&496400\\
BUF&76&29.2&32390&52.6&24.1&14.0&3.6&732.5&74400\\
CHI&722&17.0&51499&52.0&8.6&10.8&4.6&1076.5&246100\\
CIN&82&22.0&43451&55.4&36.1&14.4&5.4&839.5&127550\\
CLE&159&32.4&29435&35.7&54.6&20.3&6.0&712&60400\\
CLT&136&13.0&53785&45.3&34.4&8.5&6.2&1064&192800\\
COL&94&23.9&42040&60.7&30.9&10.1&3.6&861.5&118450\\
DEN&134&11.7&67622&82.5&4.7&5.5&3.2&1425.5&413000\\
DET&246&36.5&28990&4.9&92.1&27.9&4.0&789&41400\\
HOU&341&20.0&46495&60.3&13.4&11.2&8.0&1012&148500\\
JAX&139&14.4&51276&60.3&29.5&12.0&6.7&949&149500\\
LAX&818&17.1&55909&50.5&4.8&5.8&2.3&1573&566700\\
MIA&80&22.7&36062&87.5&5.9&11.9&5.4&1341.5&283900\\
MKE&193&25.3&37917&38.1&36.8&10.2&5.6&845&99300\\
NAS&133&14.2&56286&66.8&22.7&8.2&5.2&1065&206200\\
NYC&1757&13.8&66466&41.3&8.6&7.7&2.5&1583&603300\\
ORL&19&15.2&50202&79.7&9.2&14.8&6.1&1245&230400\\
PDX&119&11.1&69792&80.4&4.3&5.7&2.5&1355&444400\\
PHL&338&21.9&44184&39.3&31.5&11.6&5.0&1044&168700\\
PHX&253&19.7&48663&75.4&4.9&8.8&4.0&1028&187400\\
PIT&114&16.9&49411&74.1&14.3&10.7&3.8&976&119650\\
RAL&32&14.7&62150&75.0&16.1&11.2&6.4&1101.5&313050\\
SD&207&11.2&74028&70.2&4.3&5.6&2.6&1734&545700\\
SEA&126&7.9&98489&74.5&3.4&5.6&1.4&1790.5&665350\\
SFO&164&8.6&114074&49.2&2.8&6.9&0.9&2224&1098950\\
STL&101&21.3&40236&38.3&46.4&18.8&7.3&919&109400\\
WAS&153&14.7&90179&40.5&41.8&9.7&6.0&1899&545500\\
\hline
\end{tabular}
\end{center}
\caption*{N = Number of tracts; Y = tract income; Val = property value.}
\end{adjustwidth}
\end{table}

\begin{table}[ht]
    \begin{adjustwidth}{-.5in}{-.5in}  
\caption{Median Area-wide Statistics for 30 MSAs.}
        \begin{center}

\begin{tabular}{lrrrrrrrrr}

CITY&N&\%POV&MedY&\%WHT&\%BLK&\%VAC&RENTVAC&MEDRENT&MedVal\\
\hline
ATL&543&13.5&60333&42.8&34.9&8.5&5.8&1111&184900\\
AUS&278&9.5&71780&81.5&4.8&7.9&5.8&1338&272500\\
BAL&484&9.4&72118&61.6&23.2&8.4&5.6&1297&245300\\
BOS&852&7.3&88430&80.2&3.5&5.5&1.5&1568&422800\\
BUF&216&12.2&56807&87.4&4.5&8.5&1.4&809&132350\\
CHI&1610&10.8&64569&67.2&5.6&7.9&4.0&1136&224950\\
CIN&445&11.7&57619&87.8&4.7&8.4&3.4&850&143800\\
CLE&509&15.2&48456&73.5&14.3&10.0&4.5&820&110600\\
CLT&460&12.0&54721&69.8&19.4&8.5&4.9&880&159150\\
COL&373&12.1&59670&80.9&7.6&7.1&3.5&885&157800\\
DEN&333&7.5&72823&86.4&1.7&3.9&2.7&1395&341000\\
DET&956&13.4&52411&73.2&12.3&8.0&2.8&927&127250\\
HOU&828&14.9&55036&68.3&11.9&9.6&7.2&1050&152300\\
JAX&213&13.1&54349&70.3&16.4&12.2&5.4&981&164200\\
LAX&2396&13.1&66409&53.0&3.0&4.8&2.1&1610&544550\\
MIA&976&13.5&52706&79.0&9.6&11.8&5.4&1395&244100\\
MKE&387&12.4&55767&80.4&5.5&6.2&4.2&889&158900\\
NAS&308&10.8&60608&80.6&10.7&7.5&4.5&985&203850\\
NYC&3337&10.9&72833&57.9&6.9&7.4&2.5&1576&467200\\
ORL&410&13.1&53052&75.9&10.2&12.7&6.9&1098&183200\\
PDX&432&9.6&71832&82.9&1.6&4.9&1.7&1244&349650\\
PHL&1159&9.6&68207&75.0&11.0&7.7&4.2&1162&222600\\
PHX&676&13.4&56099&79.1&4.0&9.4&4.5&1080&206250\\
PIT&631&10.1&55699&92.1&2.9&9.9&4.2&827&129000\\
RAL&187&8.1&69863&70.2&15.9&7.3&4.4&1073&239000\\
SD&524&10.7&75355&75.1&3.2&5.5&2.3&1654&501650\\
SEA&607&7.6&84918&70.2&3.8&5.3&1.0&1449&406800\\
SFO&792&7.5&102092&48.0&4.1&4.9&1.4&2062&824450\\
STL&540&10.6&59239&84.5&7.1&9.3&5.1&895&152050\\
WAS&857&7.7&92656&54.0&19.1&6.0&3.7&1699&386700\\
\hline
\end{tabular}
\end{center}
\caption*{N = number of tracts; Y = tract income; Val = property value.}
\end{adjustwidth}
\end{table}

\begin{table}[ht]
    \begin{adjustwidth}{-1in}{-1in}  
\footnotesize
\caption{Correlations for All Variables; City of Milwaukee.}
        \begin{center}
\begin{tabular}{lrrrrrrrrrr}

&RPV&\%POV&lnMedY&\%BLK&PERCWHT&\%VAC&RENTVAC&lnMedVal&MEDRENT&lnMedYr\\
\hline
RPV&1&0.578&-0.650&0.624&-0.678&0.590&0.206&-0.894&-0.217&-0.232\\
PERCPOV&&1&-0.881&0.439&-0.571&0.615&0.368&-0.636&-0.343&-0.316\\
lnMedY&&&1&-0.557&0.666&-0.615&-0.356&0.742&0.453&0.352\\
PERCBLK&&&&1&-0.921&0.498&0.152&-0.641&-0.299&-0.033\\
PERCWHT&&&&&1&-0.545&-0.223&0.768&0.425&0.120\\
PERCVAC&&&&&&1&0.657&-0.550&-0.206&-0.363\\
RENTVAC&&&&&&&1&-0.257&-0.195&-0.269\\
lnMedVal&&&&&&&&1&0.567&0.294\\
MEDRENT&&&&&&&&&1&0.400\\
lnMedYr&&&&&&&&&&1\\
\hline
\end{tabular}
\end{center}
\caption*{City tracts only.}
\end{adjustwidth}
\end{table}

\begin{table}[ht]
\caption{Correlations Between RPV and Other Variables for 30 Core Cities.}
    \begin{adjustwidth}{-.75in}{-.5in}  
        \begin{center}
\begin{tabular}{lrrrrrrrrr}

&\%POV&lnMedY&\%BLK&\%WHT&\%VAC&RENTVAC&lnMedVal&MEDRENT&lnMedYr\\
\hline
ATL&0.553&-0.650&0.624&-0.650&0.481&0.049&-0.846&-0.335&-0.088\\
AUS&0.227&-0.371&0.086&-0.335&-0.108&-0.141&-0.821&0.003&0.414\\
BAL&0.297&-0.353&0.292&-0.320&0.454&-0.033&-0.722&-0.038&-0.120\\
BOS&-0.197&0.235&-0.192&0.223&-0.093&0.095&-0.448&0.553&0.025\\
BUF&0.510&-0.617&0.478&-0.524&0.449&-0.085&-0.913&-0.369&-0.218\\
CHI&0.344&-0.398&0.503&-0.494&0.438&0.122&-0.747&-0.119&-0.018\\
CIN&0.476&-0.427&0.164&-0.20&0.310&0.128&-0.836&-0.105&-0.043\\
CLE&0.204&-0.288&0.330&-0.334&0.404&0.128&-0.766&0.106&-0.184\\
CLT&0.399&-0.522&0.503&-0.537&0.057&-0.084&-0.797&-0.302&-0.004\\
COL&0.503&-0.674&0.498&-0.505&0.267&-0.071&-0.898&-0.392&0.013\\
DEN&-0.020&-0.304&0.091&-0.157&-0.264&-0.203&-0.749&0.179&0.249\\
DET&0.226&-0.215&0.239&-0.239&0.437&-0.119&-0.828&0.078&-0.121\\
HOU&0.380&-0.445&0.307&-0.305&0.060&-0.127&-0.775&-0.262&-0.098\\
JAX&0.660&-0.668&0.646&-0.640&0.360&-0.048&-0.884&-0.277&-0.287\\
LAX&0.121&-0.141&0.045&-0.111&-0.152&-0.017&-0.703&0.101&0.355\\
MIA&0.239&-0.273&-0.007&0.007&-0.001&0.118&-0.691&0.050&0.262\\
MKE&0.578&-0.65&0.624&-0.678&0.590&0.206&-0.894&-0.217&-0.232\\
NAS&-0.039&-0.227&0.148&-0.198&-0.155&-0.162&-0.830&0.114&0.271\\
NYC&-0.072&-0.019&0.064&-0.084&-0.081&-0.025&-0.678&0.051&0.092\\
ORL&0.169&-0.397&0.419&-0.493&0.007&-0.157&-0.910&-0.089&0.407\\
PDX&0.348&-0.356&0.211&-0.527&0.086&-0.154&-0.704&0.120&0.364\\
PHL&0.566&-0.58&0.421&-0.513&0.352&-0.007&-0.836&-0.295&-0.287\\
PHX&0.288&-0.294&0.130&-0.274&0.150&0.161&-0.706&-0.086&0.049\\
PIT&0.297&-0.454&0.393&-0.359&0.542&0.223&-0.810&-0.208&-0.104\\
RAL&0.256&-0.371&0.383&-0.454&0.099&-0.101&-0.774&0.070&0.345\\
SD&0.178&-0.245&0.185&-0.306&0.074&0.122&-0.667&0.241&0.427\\
SEA&0.067&-0.246&0.053&-0.140&0.156&0.167&-0.658&0.413&0.377\\
SFO&-0.176&0.070&-0.091&-0.066&0.030&0.078&-0.429&0.547&0.209\\
STL&0.544&-0.652&0.650&-0.647&0.649&0.095&-0.901&-0.398&-0.189\\
WAS&0.136&-0.11&-0.092&0.101&0.044&0.078&-0.466&0.344&0.258\\
\hline
\end{tabular}
\end{center}
\caption*{Calculated for all tracts within each city.\\
lnMedYr = log(Median year structure built).}
\end{adjustwidth}
\end{table}

\begin{table}[ht]
\caption{Summary Statistics for the  RPV Ratio.}
    \begin{adjustwidth}{-.75in}{-.5in}  
        \begin{center}
\begin{tabular}{lrrrrrrrrrrrr}

&City&&&&&&MSA&&&&&\\&Mean&SD&50\%&75\%&90\%&95\%&Mean&SD&50\%&75\%&90\%&95\%\\
\hline
ATL&7.368&3.622&6.789&9.58&11.812&13.007&7.65&4.303&7.209&9.403&11.893&13.999\\
AUS&5.542&2.492&4.99&6.513&8.161&9.222&6.141&2.944&5.735&7.146&9.22&10.541\\
BAL&10.307&6.05&8.611&12.224&16.564&20.827&7.655&4.461&6.78&8.729&12.235&15.868\\
BOS&3.897&1.345&3.912&4.569&5.296&6.028&4.482&1.407&4.433&5.314&6.224&6.896\\
BUF&10.908&5.718&11.069&14.091&19.62&21.192&8.777&4.097&7.759&9.915&13.796&17.578\\
CHI&6.094&2.495&5.62&7.182&9.271&10.81&6.793&3.527&6.227&8.026&10.363&12.049\\
CIN&8.377&4.207&7.882&10.705&14.55&16.362&7.859&3.51&7.137&9.429&12.525&14.595\\
CLE&14.229&6.293&13.907&17.223&21.926&23.57&10.41&5.412&9.191&13.483&17.25&19.807\\
CLT&7.115&4.021&6.93&9.173&10.446&11.631&6.993&4.609&6.391&8.409&10.221&10.942\\
COL&9.727&4.124&8.669&12.92&15.194&16.567&7.79&3.319&6.964&9.317&12.37&14.117\\
DEN&4.563&1.536&4.426&5.497&6.645&7.613&5.547&4.385&4.976&6.119&7.187&8.568\\
DET&23.633&11.889&22.562&29.73&37.215&43.542&13.145&10.741&9.148&17.125&25.876&32.437\\
HOU&8.83&5.807&8.291&11.349&14.561&16.149&8.84&4.634&8.316&10.863&13.493&15.009\\
JAX&8.859&3.818&8.033&10.449&14.815&16.449&7.88&3.622&7.171&9.418&12.443&15.866\\
LAX&3.448&1.096&3.362&4.034&4.755&5.286&3.702&2.139&3.567&4.174&4.873&5.473\\
MIA&6.319&2.447&5.946&7.239&9.277&10.284&7.911&5.252&6.927&9.102&12.3&15.339\\
MKE&11.149&4.761&9.935&14.013&17.295&20.395&8.391&4.484&7.014&10.231&14.975&17.293\\
NAS&6.046&1.993&6.068&7.454&8.675&9.389&5.764&1.817&5.649&6.915&8.271&9.086\\
NYC&3.504&3.049&3.097&3.929&5.059&6.289&4.481&2.841&4.006&5.497&7.079&8.328\\
ORL&9.32&5.022&6.644&12.2&15.968&18.165&7.821&4.043&6.905&8.9&12.198&14.221\\
PDX&4.009&0.926&3.94&4.487&5.221&5.541&4.344&1.617&4.187&4.959&5.78&6.237\\
PHL&9.062&4.746&7.577&10.922&16.187&18.031&7.622&3.939&6.642&8.806&12.59&15.815\\
PHX&6.979&5.115&6.285&8.088&9.428&11.641&7.543&7.262&6.142&8.008&10.092&13.724\\
PIT&10.432&5.612&9.535&12.872&18.143&21.933&9.14&4.923&7.935&10.609&14.71&19.274\\
RAL&4.733&1.948&4.521&5.794&6.364&7.416&5.386&1.714&5.232&6.399&7.738&8.4\\
SD&4.028&1.287&3.904&4.595&5.555&6.421&4.435&5.162&3.83&4.668&5.549&6.825\\
SEA&3.435&0.989&3.221&3.899&4.628&5.437&4.411&2.562&4.211&5.168&6.16&6.697\\
SFO&2.427&0.749&2.355&2.772&3.321&3.971&3.286&1.29&3.086&3.876&4.852&5.605\\
STL&11.309&5.958&9.849&15.086&18.92&22.619&8.892&5.809&7.134&10.887&15.967&18.952\\
WAS&4.276&1.569&4.148&5.131&6.207&7.162&5.368&2.156&5.005&6.606&7.799&8.83\\

\hline
\end{tabular}
\end{center}
\end{adjustwidth}
\end{table}

\begin{table}[ht]
    \begin{adjustwidth}{-1in}{-.5in}  
\footnotesize
\caption{Spatial Regression Results, Core-City Tracts. DV = RPV Z-Score.}
        \begin{center}
\begin{tabular}{lrrrrrrrrr}

&Const&lnMedY&lnMedVal&PERCVAC&PERCWHT&PERCBLK&PERCRENT&rho&AIC\\
\hline
ATL&19.511 (7.46)&0.692 (3.06)&-2.273 (13.92)&-0.002 (0.28)&0.013 (3.31)&&0.007 (2.07)&-0.056&-61.9\\
&21.123 (6.89)&0.586 (2.37)&-2.203 (13.98)&-0.002 (0.36)&&-0.012 (3.21)&0.005 (1.24)&-0.054&-62.3\\
AUS&21.706 (10.56)&0.787 (5.03)&-2.532 (15.22)&0.052 (4.26)&0.011 (2.94)&&0.007 (2.59)&-0.015&-84.9\\
&22.233 (10.93)&0.757 (4.90)&-2.463 (15.58)&0.052 (4.30)&&-0.021 (3.61)&0.006 (2.23)&-0.027&-82.9\\
BAL&13.096 (6.02)&0.789 (3.71)&-1.884 (12.65)&0.012 (2.45)&0.007 (2.80)&&0.006 (1.67)&-0.122&-160.0\\
&13.813 (6.08)&0.767 (3.61)&-1.870 (12.65)&0.013 (2.50)&&-0.006 (3.00)&0.006 (1.47)&-0.123&-159.4\\
BOS&24.145 (7.18)&1.036 (5.16)&-2.851 (11.88)&0.034 (2.64)&0.012 (4.11)&&0.015 (3.74)&0.166&-142.7\\
&22.000 (6.78)&1.214 (6.57)&-2.765 (11.51)&0.033 (2.56)&&-0.010 (3.53)&0.015 (3.44)&0.196&-145.5\\
BUF&9.604 (4.56)&0.380 (1.58)&-1.218 (8.35)&0.019 (2.64)&-0.002 (1.00)&&0.003 (0.60)&0.146&-35.1\\
&9.718 (4.72)&0.367 (1.55)&-1.233 (8.50)&0.019 (2.62)&&0.002 (1.08)&0.004 (0.73)&0.135&-34.8\\
CHI&10.089 (11.04)&1.149 (13.85)&-1.871 (22.46)&0.010 (3.05)&-0.002 (1.41)&&0.010 (6.42)&0.205&-627.2\\
&9.676 (11.57)&1.187 (15.00)&-1.882 (23.18)&0.007 (2.07)&&0.003 (4.02)&0.010 (6.44)&0.167&-616.2\\
CIN&12.820 (6.40)&0.778 (4.03)&-1.919 (15.55)&0.007 (1.24)&0.009 (3.45)&&0.017 (4.43)&-0.086&-44.7\\
&14.325 (7.01)&0.740 (4.03)&-1.937 (16.19)&0.009 (1.53)&&-0.010 (4.23)&0.016 (4.25)&-0.114&-42.6\\
CLE&11.844 (6.34)&0.688 (3.72)&-1.807 (13.37)&0.011 (1.80)&0.002 (0.88)&&0.015 (3.81)&-0.097&-142.2\\
&11.920 (6.21)&0.693 (3.77)&-1.805 (13.37)&0.011 (1.77)&&-0.001 (0.78)&0.014 (3.75)&-0.103&-142.4\\
CLT&12.689 (6.40)&1.318 (6.97)&-2.326 (15.90)&0.021 (2.12)&0.008 (2.72)&&0.014 (5.05)&-0.119&-89.0\\
&10.046 (4.59)&1.481 (7.46)&-2.218 (15.28)&0.019 (1.84)&&-0.002 (0.67)&0.015 (5.05)&-0.168&-93.2\\
COL&17.846 (9.24)&0.343 (1.46)&-1.888 (11.54)&-0.005 (0.90)&0.006 (2.61)&&0.006 (2.14)&0.067&-48.2\\
&17.406 (8.78)&0.407 (1.74)&-1.871 (11.34)&-0.005 (0.80)&&-0.004 (2.02)&0.007 (2.15)&0.071&-49.5\\
DEN&20.164 (9.03)&1.909 (8.52)&-3.319 (15.27)&0.022 (1.59)&0.005 (1.30)&&0.017 (5.26)&0.079&-100.5\\
&19.774 (8.64)&1.971 (8.78)&-3.312 (15.11)&0.022 (1.60)&&-0.002 (0.37)&0.017 (5.31)&0.069&-101.1\\
DET&10.770 (7.62)&0.292 (2.26)&-1.330 (17.37)&0.009 (3.27)&-0.001 (0.37)&&0.006 (2.20)&-0.026&-198.8\\
&10.694 (7.51)&0.292 (2.26)&-1.330 (17.39)&0.009 (3.22)&&0.001 (0.59)&0.006 (2.18)&-0.029&-198.7\\
HOU&5.075 (5.48)&1.213 (11.03)&-1.586 (22.39)&0.004 (0.73)&0.001 (0.41)&&0.015 (8.33)&0.030&-268.5\\
&4.874 (5.11)&1.228 (11.24)&-1.580 (22.31)&0.003 (0.62)&&0.000 (0.28)&0.015 (8.30)&0.029&-268.6\\
JAX&12.031 (6.97)&1.447 (6.36)&-2.384 (15.11)&0.009 (1.42)&0.002 (0.70)&&0.010 (3.70)&0.122&-72.5\\
&12.269 (6.44)&1.430 (6.12)&-2.376 (15.20)&0.009 (1.40)&&-0.001 (0.65)&0.009 (3.61)&0.124&-72.5\\
LAX&18.075 (16.00)&1.408 (15.73)&-2.605 (29.68)&0.036 (7.12)&0.003 (2.13)&&0.011 (8.40)&0.163&-714.3\\
&17.266 (15.65)&1.465 (16.62)&-2.579 (29.72)&0.036 (7.12)&&0.000 (0.21)&0.011 (8.07)&0.170&-717.6\\
MIA&18.682 (7.99)&0.936 (3.98)&-2.434 (13.70)&0.025 (3.09)&0.012 (4.82)&&0.010 (2.50)&-0.069&-53.8\\
&20.398 (8.20)&0.883 (3.69)&-2.430 (13.72)&0.025 (3.08)&&-0.013 (4.86)&0.009 (2.30)&-0.061&-53.5\\
MKE&13.159 (8.82)&0.744 (5.42)&-1.922 (16.01)&0.011 (2.19)&0.004 (2.92)&&0.014 (6.14)&0.247&-93.7\\
&11.835 (8.00)&0.783 (5.59)&-1.827 (15.45)&0.011 (2.13)&&0.001 (0.56)&0.013 (5.83)&0.156&-93.1\\
NAS&18.292 (7.57)&1.205 (6.61)&-2.633 (15.90)&0.009 (1.14)&0.005 (2.12)&&0.013 (4.95)&-0.002&-81.1\\
&18.891 (7.47)&1.185 (6.43)&-2.626 (16.03)&0.009 (1.20)&&-0.005 (2.24)&0.013 (4.75)&-0.006&-80.8\\
NYC&14.442 (21.03)&0.775 (14.70)&-1.782 (38.09)&0.010 (3.48)&0.003 (3.95)&&0.008 (9.38)&0.104&-1,785.6\\
&14.168 (21.33)&0.814 (16.23)&-1.782 (38.09)&0.010 (3.59)&&-0.002 (3.90)&0.008 (9.44)&0.097&-1,784.8\\
ORL&11.901 (4.29)&1.219 (5.13)&-2.171 (12.25)&-0.013 (1.56)&0.005 (1.43)&&0.015 (3.48)&0.020&0.5\\
&12.129 (3.93)&1.194 (4.66)&-2.126 (12.09)&-0.013 (1.57)&&-0.004 (1.00)&0.014 (3.27)&0.019&0.0\\
PDX&25.864 (6.09)&0.778 (2.48)&-2.634 (7.68)&0.044 (2.09)&-0.014 (1.43)&&0.011 (2.10)&-0.044&-120.8\\
&27.904 (7.32)&0.671 (2.24)&-2.789 (8.70)&0.045 (2.11)&&0.013 (1.05)&0.010 (1.95)&-0.017&-121.2\\
PHL&9.672 (8.87)&0.696 (6.79)&-1.493 (16.09)&0.011 (2.27)&0.003 (2.07)&&0.012 (6.25)&0.234&-242.6\\
&9.632 (9.08)&0.712 (7.08)&-1.487 (16.31)&0.012 (2.49)&&-0.002 (2.31)&0.012 (6.12)&0.230&-241.8\\
PHX&6.163 (3.73)&1.898 (10.59)&-2.247 (20.13)&0.028 (4.19)&-0.002 (0.74)&&0.011 (4.28)&-0.208&-219.4\\
&6.163 (3.72)&1.896 (10.56)&-2.262 (20.50)&0.028 (4.13)&&0.001 (0.22)&0.012 (4.36)&-0.211&-219.7\\
PIT&10.828 (5.81)&0.325 (1.45)&-1.296 (8.12)&0.030 (3.49)&0.003 (1.13)&&0.007 (1.70)&0.003&-92.1\\
&11.964 (6.15)&0.283 (1.30)&-1.328 (8.33)&0.032 (3.70)&&-0.005 (2.09)&0.007 (1.75)&0.014&-90.6\\
RAL&17.332 (4.16)&1.743 (4.43)&-3.000 (11.57)&0.044 (3.55)&0.004 (0.70)&&0.012 (2.02)&-0.296&-16.8\\
&18.013 (4.24)&1.704 (4.33)&-2.990 (12.04)&0.045 (3.62)&&-0.004 (0.90)&0.011 (1.93)&-0.310&-16.7\\
SD&17.567 (8.55)&2.146 (11.83)&-3.241 (17.23)&0.034 (3.45)&0.000 (0.02)&&0.016 (5.96)&0.047&-174.3\\
&18.725 (8.53)&2.103 (11.65)&-3.287 (18.39)&0.035 (3.56)&&-0.010 (1.31)&0.016 (5.88)&0.049&-173.4\\
SEA&27.493 (8.31)&1.646 (6.94)&-3.632 (13.87)&0.023 (1.50)&0.014 (3.95)&&0.023 (6.94)&0.067&-101.7\\
&26.087 (7.96)&1.746 (7.38)&-3.524 (13.62)&0.020 (1.28)&&-0.023 (3.35)&0.023 (6.72)&0.053&-103.6\\
SFO&36.739 (7.20)&0.750 (3.05)&-3.305 (9.56)&0.009 (0.63)&0.018 (2.83)&&-0.006 (1.45)&0.135&-185.8\\
&31.282 (8.66)&1.079 (5.92)&-3.149 (9.99)&0.011 (0.77)&&-0.030 (3.54)&0.001 (0.45)&0.132&-183.6\\
STL&13.450 (6.34)&0.290 (1.38)&-1.419 (10.03)&-0.003 (0.55)&-0.002 (0.90)&&0.003 (0.75)&0.080&-57.5\\
&13.418 (6.11)&0.277 (1.31)&-1.422 (10.02)&-0.003 (0.48)&&0.002 (0.66)&0.003 (0.87)&0.087&-57.7\\
WAS&31.088 (11.85)&0.752 (3.77)&-3.115 (17.26)&-0.010 (1.17)&0.026 (8.56)&&0.009 (2.91)&0.073&-118.1\\
&36.109 (12.11)&0.621 (3.05)&-3.198 (17.86)&-0.006 (0.67)&&-0.026 (9.00)&0.005 (1.67)&0.085&-116.1\\
\hline
\end{tabular}
\caption*{t-statistics in parentheses. \\AIC = Akaike Information Criterion for Goodness of Fit.}
\end{center}

\end{adjustwidth}
\end{table}

\begin{table}[ht]
    \begin{adjustwidth}{-1in}{-.5in}  
\footnotesize

\caption{Spatial Regression Results, MSA Tracts. DV = RPV Z-Score.}
        \begin{center}
\begin{tabular}{lrrrrrrrrr}

&Const&lnMedY&lnMedVal&PERCVAC&PERCWHT&PERCBLK&PERCRENT&rho&AIC\\
\hline
ATL&11.537 (9.69)&1.218 (9.16)&-2.139 (25.45)&0.017 (4.50)&0.008 (5.55)&&0.013 (8.13)&-0.050&-443.7\\
&11.504 (9.18)&1.238 (9.18)&-2.105 (25.02)&0.018 (4.83)&&-0.005 (4.54)&0.012 (7.25)&-0.071&-449.0\\
AUS&16.290 (12.48)&0.839 (6.42)&-2.112 (20.42)&0.006 (0.88)&0.001 (0.28)&&0.015 (7.98)&-0.183&-211.3\\
&16.761 (12.55)&0.811 (6.17)&-2.115 (20.59)&0.006 (0.86)&&-0.007 (1.50)&0.015 (8.18)&-0.177&-209.9\\
BAL&11.397 (9.49)&1.088 (8.52)&-1.953 (19.11)&0.016 (4.38)&0.002 (1.43)&&0.007 (3.77)&-0.034&-416.6\\
&12.233 (9.68)&1.055 (8.24)&-1.976 (19.40)&0.017 (4.59)&&-0.003 (2.55)&0.007 (3.92)&-0.032&-414.3\\
BOS&15.502 (14.99)&1.433 (13.27)&-2.514 (25.90)&-0.002 (0.59)&0.002 (1.52)&&0.019 (12.18)&0.032&-815.7\\
&15.761 (14.62)&1.429 (13.08)&-2.515 (25.95)&-0.002 (0.45)&&-0.002 (1.47)&0.018 (12.08)&0.030&-815.7\\
BUF&12.757 (7.89)&0.417 (2.20)&-1.443 (11.54)&0.013 (2.10)&-0.006 (3.64)&&0.000 (0.11)&0.146&-131.2\\
&13.271 (7.99)&0.315 (1.64)&-1.444 (11.25)&0.015 (2.33)&&0.003 (1.54)&0.002 (0.72)&0.177&-137.4\\
CHI&10.050 (15.83)&1.195 (18.38)&-1.924 (34.86)&0.006 (2.28)&0.000 (0.57)&&0.011 (10.41)&0.162&-1,470.0\\
&9.698 (15.47)&1.225 (19.19)&-1.926 (35.07)&0.004 (1.40)&&0.002 (2.36)&0.011 (10.59)&0.153&-1,464.9\\
CIN&11.730 (10.78)&1.264 (10.34)&-2.203 (26.42)&0.004 (0.90)&-0.001 (0.75)&&0.017 (8.67)&0.063&-321.1\\
&11.845 (10.74)&1.236 (10.03)&-2.195 (26.30)&0.005 (1.02)&&-0.001 (0.73)&0.018 (9.18)&0.072&-321.5\\
CLE&11.459 (11.70)&0.836 (8.14)&-1.798 (22.88)&0.009 (2.93)&-0.001 (0.99)&&0.008 (4.75)&-0.059&-354.8\\
&11.373 (11.44)&0.837 (8.14)&-1.800 (22.93)&0.009 (2.94)&&0.001 (0.87)&0.008 (4.93)&-0.055&-354.8\\
CLT&5.850 (4.89)&1.758 (10.11)&-2.149 (17.45)&-0.001 (0.19)&-0.001 (0.35)&&0.022 (10.04)&-0.088&-470.4\\
&5.743 (4.53)&1.765 (10.05)&-2.151 (17.54)&-0.001 (0.22)&&0.001 (0.38)&0.023 (10.74)&-0.088&-470.4\\
COL&11.219 (10.09)&1.111 (7.58)&-1.997 (18.22)&0.007 (1.59)&-0.002 (1.55)&&0.012 (7.08)&0.211&-245.0\\
&11.034 (9.88)&1.124 (7.69)&-2.010 (18.32)&0.007 (1.61)&&0.001 (0.63)&0.013 (7.73)&0.226&-247.4\\
DEN&9.754 (5.71)&2.172 (11.07)&-2.790 (21.33)&0.015 (2.25)&0.008 (2.70)&&0.019 (7.69)&-0.047&-306.6\\
&10.580 (6.20)&2.151 (11.04)&-2.780 (21.56)&0.016 (2.38)&&-0.016 (3.24)&0.018 (7.81)&-0.055&-305.2\\
DET&6.049 (9.60)&0.857 (12.61)&-1.352 (32.64)&0.015 (7.90)&0.000 (0.20)&&0.005 (4.39)&-0.041&-653.7\\
&6.049 (9.52)&0.858 (12.62)&-1.354 (32.68)&0.015 (7.96)&&0.000 (0.04)&0.005 (4.64)&-0.039&-653.6\\
HOU&7.703 (11.04)&1.304 (15.46)&-1.891 (33.23)&0.004 (1.60)&-0.001 (1.41)&&0.016 (13.37)&-0.006&-677.0\\
&7.765 (10.76)&1.289 (15.26)&-1.890 (33.07)&0.004 (1.44)&&0.000 (0.10)&0.017 (13.78)&-0.005&-678.0\\
JAX&10.975 (7.49)&1.074 (5.59)&-1.924 (15.06)&0.017 (5.07)&-0.002 (0.96)&&0.008 (3.69)&0.098&-126.8\\
&10.584 (6.74)&1.097 (5.64)&-1.928 (15.21)&0.017 (4.92)&&0.002 (1.08)&0.009 (3.90)&0.093&-126.6\\
LAX&14.587 (19.02)&1.308 (17.65)&-2.263 (38.83)&0.026 (6.77)&0.002 (2.34)&&0.011 (10.33)&-0.103&-2,705.6\\
&14.467 (19.14)&1.330 (18.37)&-2.264 (38.90)&0.027 (7.07)&&-0.003 (2.02)&0.011 (10.32)&-0.105&-2,706.6\\
MIA&12.685 (18.95)&0.962 (12.50)&-1.936 (36.40)&0.012 (7.95)&0.005 (6.72)&&0.007 (5.76)&-0.013&-832.7\\
&13.418 (19.47)&0.937 (12.16)&-1.931 (36.46)&0.012 (8.02)&&-0.006 (7.08)&0.006 (5.49)&-0.012&-830.4\\
MKE&9.222 (8.37)&0.719 (6.18)&-1.469 (15.61)&0.019 (4.61)&0.000 (0.06)&&0.007 (4.29)&0.249&-180.3\\
&8.823 (8.11)&0.727 (6.29)&-1.446 (15.92)&0.018 (4.35)&&0.003 (2.54)&0.007 (4.34)&0.193&-173.2\\
NAS&9.142 (6.00)&1.963 (11.20)&-2.528 (19.00)&0.006 (0.90)&-0.007 (3.40)&&0.021 (9.08)&0.055&-260.8\\
&8.413 (5.31)&1.967 (10.97)&-2.526 (18.80)&0.005 (0.79)&&0.006 (2.74)&0.022 (9.92)&0.066&-262.9\\
NYC&12.927 (28.39)&0.757 (19.81)&-1.671 (50.89)&0.004 (2.77)&0.001 (1.51)&&0.007 (11.68)&0.048&-3,179.1\\
&13.228 (28.65)&0.748 (20.08)&-1.681 (51.07)&0.004 (2.98)&&-0.002 (3.47)&0.007 (11.78)&0.046&-3,173.9\\
ORL&11.357 (9.67)&1.481 (9.87)&-2.352 (24.06)&0.004 (1.37)&0.005 (3.18)&&0.014 (8.43)&-0.073&-337.6\\
&12.105 (9.80)&1.447 (9.56)&-2.342 (24.00)&0.005 (1.52)&&-0.005 (2.92)&0.014 (8.23)&-0.080&-338.5\\
PDX&18.986 (10.61)&1.428 (8.20)&-2.764 (18.58)&0.001 (0.21)&-0.004 (1.10)&&0.019 (7.53)&-0.015&-436.8\\
&18.446 (10.47)&1.504 (8.56)&-2.819 (18.96)&0.001 (0.18)&&0.019 (2.22)&0.019 (7.94)&-0.012&-434.9\\
PHL&11.212 (16.34)&0.742 (11.05)&-1.625 (28.77)&0.012 (3.97)&0.000 (0.24)&&0.010 (9.16)&0.170&-853.1\\
&11.674 (16.64)&0.721 (10.83)&-1.641 (29.08)&0.013 (4.34)&&-0.001 (1.93)&0.010 (9.26)&0.172&-851.9\\
PHX&4.620 (5.01)&1.653 (15.59)&-1.956 (31.29)&0.022 (6.48)&0.006 (2.71)&&0.011 (7.52)&-0.039&-578.0\\
&4.764 (5.14)&1.657 (15.54)&-1.933 (31.01)&0.025 (7.59)&&-0.001 (0.13)&0.010 (6.57)&-0.034&-581.5\\
PIT&8.034 (8.88)&0.958 (8.57)&-1.555 (20.03)&0.014 (3.80)&-0.009 (6.34)&&0.010 (6.36)&0.102&-471.6\\
&6.966 (7.57)&0.961 (8.50)&-1.542 (19.64)&0.014 (3.79)&&0.008 (5.09)&0.012 (7.52)&0.119&-479.4\\
RAL&16.146 (8.85)&1.447 (5.97)&-2.650 (13.37)&0.008 (0.87)&-0.004 (1.50)&&0.020 (5.80)&0.018&-147.5\\
&15.141 (7.51)&1.554 (6.24)&-2.697 (13.93)&0.007 (0.75)&&0.005 (1.45)&0.021 (6.66)&0.017&-147.6\\
SD&10.427 (7.86)&1.553 (10.83)&-2.211 (23.76)&0.017 (3.63)&0.008 (3.65)&&0.011 (5.77)&-0.070&-525.0\\
&11.405 (8.28)&1.459 (10.04)&-2.157 (23.31)&0.020 (4.26)&&-0.014 (2.24)&0.010 (5.36)&-0.073&-529.2\\
SEA&9.019 (6.63)&1.971 (11.92)&-2.468 (21.39)&0.003 (0.39)&-0.003 (1.62)&&0.020 (9.79)&-0.128&-641.7\\
&8.893 (6.20)&1.961 (11.59)&-2.468 (21.33)&0.002 (0.20)&&0.002 (0.35)&0.021 (10.43)&-0.125&-642.7\\
SFO&18.853 (17.71)&1.144 (13.71)&-2.392 (26.48)&-0.007 (1.43)&0.002 (1.65)&&0.008 (7.15)&0.058&-682.3\\
&21.083 (19.48)&1.053 (12.87)&-2.466 (27.76)&-0.005 (1.06)&&-0.014 (5.89)&0.009 (7.44)&0.049&-666.4\\
STL&9.807 (8.20)&0.981 (6.82)&-1.718 (19.36)&0.000 (0.02)&-0.006 (4.68)&&0.007 (3.41)&-0.044&-442.9\\
&8.987 (7.40)&1.009 (6.97)&-1.727 (19.35)&0.000 (0.04)&&0.005 (3.99)&0.009 (4.25)&-0.035&-445.6\\
WAS&13.741 (16.40)&1.459 (17.35)&-2.423 (34.79)&0.003 (0.72)&0.000 (0.23)&&0.018 (17.23)&0.096&-690.2\\
&14.317 (16.39)&1.415 (16.59)&-2.424 (34.87)&0.004 (0.95)&&-0.001 (1.59)&0.018 (17.21)&0.104&-689.9\\
\hline

\end{tabular}
\caption*{t-statistics in parentheses. \\AIC = Akaike Information Criterion for Goodness of Fit.}
\end{center}
\end{adjustwidth}

\end{table}

\nocite{Bayeretal}
\nocite{Colburn}
\nocite{DesmondWilmers}
\nocite{Desmond}
\nocite{DingKnapp}
\nocite{Early}
\nocite{GretherM}
\nocite{Gilderb}
\nocite{Haugen}
\nocite{Ihlanf}
\nocite{LeSage}
\nocite{Sen}
\nocite{Zeitz}
\bibliographystyle{abbrv}
\bibliography{RPVpaper}

\begin{thebibliography}{10}

\bibitem{Bayeretal}
P.~Bayer, M.~Casey, F.~Ferreira, and R.~McMillan.
\newblock {Racial and ethnic price differentials in the housing market}.
\newblock {\em Journal of Urban Economics}, 102(C):91--105, 2017.

\bibitem{Colburn}
G.~Colburn and R.~Allen.
\newblock {Rent burden and the Great Recession in the USA}.
\newblock {\em Urban Studies}, 55(1):226--243, 2018.

\bibitem{Desmond}
M.~Desmond.
\newblock {\em { Evicted: Poverty and Profit in the American City}}.
\newblock Crown, New York, 2019.

\bibitem{DesmondWilmers}
M.~Desmond and N.~Wilmers.
\newblock {Do the Poor Pay More for Housing? Exploitation, Profit, and Risk in
  Rental Markets}.
\newblock {\em American Journal of Sociology}, 124(4):1090--1124, 2019.

\bibitem{DingKnapp}
C.~Ding and G.~Knaap.
\newblock Property values in inner‐city neighborhoods: The effects of
  homeownership, housing investment, and economic development.
\newblock {\em Housing Policy Debate}, 13(4):701--727, 2002.

\bibitem{Early}
D.~W. Early, P.~E. Carrillo, and E.~O. Olsen.
\newblock Racial rent differences in u.s. housing markets: Evidence from the
  housing voucher program.
\newblock {\em Journal of Regional Science}, 59(4):669--700, 2019.

\bibitem{Gilderb}
J.~I. Gilderbloom and R.~P. Appelbaum.
\newblock Toward a sociology of rent: Are rental housing markets competitive?
\newblock {\em Social Problems}, 34(3):261--276, 1987.

\bibitem{GretherM}
D.~Grether and P.~Mieszkowski.
\newblock Determinants of real estate values.
\newblock {\em Journal of Urban Economics}, 1(2):127--145, 1974.

\bibitem{Haugen}
R.~A. Haugen and A.~J. Heins.
\newblock {A Market Separation Theory of Rent Differentials in Metropolitan
  Areas}.
\newblock {\em The Quarterly Journal of Economics}, 83(4):660--672, 1969.

\bibitem{Ihlanf}
K.~Ihlanfeldt and T.~Mayock.
\newblock Price discrimination in the housing market.
\newblock {\em Journal of Urban Economics}, 66(2):125--140, 2009.

\bibitem{LeSage}
J.~P. LeSage.
\newblock {What Regional Scientists Need to Know about Spatial Econometrics}.
\newblock {\em The Review of Regional Studies}, 44(1):13--32, Spring 2014.

\bibitem{Sen}
P.~K. Sen.
\newblock Estimates of the regression coefficient based on kendall's tau.
\newblock {\em Journal of the American Statistical Association},
  63(324):1379--1389, 1968.

\bibitem{Zeitz}
E.~Zietz and S.~Sirmans.
\newblock An exploration of inner-city property markets.
\newblock {\em Journal of Real Estate Literature}, 12(3):323--360, 2004.

\end{thebibliography}

\end{document}